\documentclass[oupdraft]{bio}
\usepackage{url}
\usepackage{booktabs}
\usepackage{amsthm}
\usepackage{setspace}

\def\boxit#1{\vbox{\hrule\hbox{\vrule\kern6pt
          \vbox{\kern6pt#1\kern6pt}\kern6pt\vrule}\hrule}}

\def\sumi{\sum_{i=1}^n}

\def\wh{\widehat}

\def\log{\hbox{log}}

\def\Normal{\hbox{Normal}}

\def\bse{\begin{eqnarray*}}
\def\ese{\end{eqnarray*}}
\def\be{\begin{eqnarray}}
\def\ee{\end{eqnarray}}
\def\bq{\begin{equation}}
\def\eq{\end{equation}}
\def\bse{\begin{eqnarray*}}
\def\ese{\end{eqnarray*}}
\def\pr{\hbox{pr}}
\def\wh{\widehat}
\def\trans{^{\rm T}}

\def\b1e{{\mathbf e}}

\def\bO{{\mathbf O}}

\def\bS{{\mathbf S}}

\def\bzero{{\mathbf 0}}

\def\bPhi{{\boldsymbol{\Phi}}}
\def\bPsi{{\boldsymbol{\Psi}}}
\def\bbeta{{\boldsymbol{\beta}}}

\def\btheta{{\boldsymbol{\theta}}}

\def\boldeta{{\boldsymbol{\eta}}}

\def\bgamma{{\boldsymbol{\gamma}}}

\def\bA{{\mathbf A}}

\def\bB{{\mathbf B}}

\def\b1e{{\mathbf e}}

\def\bB{{\mathbf B}}

\def\bS{{\mathbf S}}

\def\bz{{\mathbf z}}
\def\bZ{{\bf Z}}

\def\bZ{{\mathbf Z}}
\def\bzero{{\mathbf 0}}
\def\0{{\mathbf 0}}

\def\bLambda{{\boldsymbol{\Lambda}}}

\def\bUpsilon{{\boldsymbol{\Upsilon}}}
\def\Normal{\hbox{Normal}}

\usepackage{xr-hyper}
\usepackage{makecell}
\usepackage{hyperref}
\hypersetup{
    colorlinks=true,
    allcolors  =blue,
    pdftitle={Overleaf Example},
    pdfpagemode=FullScreen,
    }

\def\sextant{\quad \quad \quad \quad}

\def\independenT#1#2{\mathrel{\rlap{$#1#2$}\mkern2mu{#1#2}}}

\newcommand{\notindependent}{\mathrel{\not\perp\!\!\!\perp}}
\newcommand{\independent}{\protect\mathpalette{\protect\independenT}{\perp}}

\def\spacingset#1{\renewcommand{\baselinestretch}%
{#1}\small\normalsize}

\newtheoremstyle{tightitalic}
  {0pt}
  {0pt}
  {\itshape}
  {}
  {\bfseries}
  {. }
  {0.5em}
  {\underline{\thmname{#1}~\thmnumber{#2}}\thmnote{ (\textit{#3})}}

\theoremstyle{tightitalic}

\newtheorem{Th}{Theorem}
\newtheorem{Prop}{Proposition}
\newtheorem{Coro}{Corollary}

\begin{document}

\title{Robust Estimation under Outcome Dependent Right Censoring in Huntington Disease: Estimators for Low and High Censoring Rates}
\author{JESUS E. VAZQUEZ$^{\ast,1}$, YANYUAN MA$^{2}$, KAREN MARDER$^{3}$, TANYA P. GARCIA$^{4}$\\[3pt]
\textit{\begin{tabular}{@{}c@{}}
$^{1}$Department of Biostatistics, Johns Hopkins University, Baltimore, MD 21231\\
$^{2}$Department of Statistics, Penn State University, State College, PA 16802\\
$^{3}$Department of Neurology, Columbia University Medical Center, New York, NY 10032\\
$^{4}$Department of Biostatistics, University of North Carolina at Chapel Hill, Chapel Hill, NC 27516
\end{tabular}}
\\[6pt]
{jvazqu18@jh.edu}
}

\markboth%
{J. E. Vazquez and others}
{Robust estimation under outcome dependent right censoring}

\maketitle
\footnotetext{To whom correspondence should be addressed.}



\vspace{-2em}

\begin{abstract}
{
Across health applications, researchers model outcomes as a function of time to an event, but the event time is right-censored for participants who exit the study or otherwise do not experience the event during follow-up. When censoring depends on the outcome---as in neurodegenerative disease studies where dropout is potentially related to disease severity---standard regression estimators produce biased estimates. We develop three consistent estimators for this outcome-dependent censoring setting: two augmented inverse probability weighted (AIPW) estimators and one maximum likelihood estimator (MLE). We establish their asymptotic properties and derive their robust sandwich variance estimators that account for nuisance parameter estimation. A key contribution is demonstrating that the choice of estimator to use depends on the censoring rate---the MLE performs best under low censoring rates, while the AIPW estimators yield lower bias and a higher nominal coverage under high censoring rates. We apply our estimators to Huntington disease data to characterize health decline leading up to mild cognitive impairment onset. The AIPW estimator with robustness matrix provided clinically-backed estimates with improved precision over inverse probability weighting, while MLE exhibited bias. Our results provide practical guidance for estimator selection based on censoring rate.}{Censored covariate; Outcome dependent; Robustness; Efficiency; Huntington Disease}
\end{abstract}

\section{Introduction}
\label{sec:chp3-introduction}

The right-censored covariate problem arises in regression analysis when the outcome is fully observed but at least one covariate is right-censored---a distinct setting from survival analysis, where the outcome is right-censored. This problem appears across various scientific disciplines. In \textit{economics}, researchers model the probability of filing unemployment insurance claims as a function of unemployment duration to understand how the length of joblessness affects benefit claiming behavior  \citep{Lvetal2017}. In \textit{Alzheimer disease} research, investigators model offspring amyloid burden as a function of maternal age at dementia onset to understand how maternal disease timing affects offspring brain pathology \citep{Atemetal2017}.  In \textit{cardiovascular health}, epidemiologists model cholesterol levels as a function of age at cardiovascular disease onset among smokers to understand how cholesterol changes before disease develops \citep{MatsouakaAtem2020}. Each analysis models an outcome as a function of time to event---but these covariates are right-censored for many participants because total unemployment duration, maternal dementia onset, and cardiovascular disease onset were unobserved at the time of data collection.

Formally, the right-censored covariate problem has the structure: $Y = m(X, \bZ; \bbeta) + \epsilon$ where $Y$ is a fully observed outcome, $m(\cdot)$ is a linear or non-linear mean function of covariates $(X,\bZ)$ indexed by parameter vector $\bbeta$, and $\epsilon\sim\Normal(0,\sigma^2)$ represents random normally distributed error. The values for $\bZ$ are always observed, whereas the covariate $X$ is right-censored: instead of observing $X$, we observe $(W,\Delta)$ where $W = \min(X, C)$ and $\Delta = I(X\leq C)$. Here, $C$ is a random censoring variable.

Most existing estimators for the right-censored covariate problem assume \textit{outcome-independent censoring}, where the censoring mechanism does not depend on the outcome once conditioning on the covariates, such as $C \independent Y \mid (X, \bZ)$ \citep{AshnerGarcia2023, lotspeich2024making, vazquez2024establishing} or $C \independent (Y,X) | \bZ$ \citep{lee2024robust, Qianetal2018}. The assumption is reasonable when censoring arises from administrative factors such as predetermined study end dates. However, in studies of progressive diseases, censoring may depend directly on the outcome, violating the outcome-independent assumption---a situation that occurs in Huntington disease research. 

The composite Unified Huntington Disease Rating Scale (cUHDRS), which integrates assessments of motor, cognitive, and functional abilities, has become a primary outcome in clinical trials evaluating therapies for Huntington disease (\citet{rodrigues2019huntington}, \href{https://www.clinicaltrials.gov/study/NCT03761849?rank=3&term=AREA%5BConditionSearch%5D(Huntington%20disease)&utm_source=chatgpt.com&tab=table#trial-description}{NCT03761849}; Hoffmann-La Roche). Understanding how cUHDRS changes in early disease stages is important for characterizing disease progression and informing trial design. Mild cognitive impairment (MCI)---which manifests as detectable cognitive changes before more severe motor or functional impairment---represents such an early stage and a potential target for preventive interventions. We model cUHDRS as a function of time to MCI onset to characterize the trajectory during this pre-MCI period. The approach aligns individuals at a common disease event rather than calendar time or study enrollment, isolating true disease progression from between-person heterogeneity in baseline disease stage. The resulting trajectory reveals when and how rapidly cUHDRS declines as individuals approach MCI, providing disease progression context for trials testing whether interventions can slow decline.

Estimating the trajectory creates a right-censored covariate problem with outcome-dependent censoring. The outcome (cUHDRS) is measured at each study visit, but the covariate of interest (time to MCI onset) is right-censored for participants who exit studies before developing MCI. Censoring occurs for two reasons: administrative censoring when the study ends before MCI is reached, and dropout due to worsening symptoms, caregiver burden, or difficulty attending study visits. The censoring mechanism is outcome-dependent because the diagnostic criteria for MCI overlap with the cognitive component of the cUHDRS, creating a direct link between cUHDRS scores and both the likelihood of reaching MCI and the likelihood of dropout. Individuals with higher cUHDRS scores at baseline---indicating better cognitive, motor, and functional status---are less likely to meet MCI diagnostic criteria during follow-up and less likely to drop out, making censoring more common in this group. On the other hand, individuals with lower cUHDRS scores are more likely to reach MCI earlier (less likely to be censored due to study end) but also more likely to drop out before MCI due to symptom severity \citep{williams2009emotional}. Under this pattern, the probability of censoring depends directly on cUHDRS values, creating dependence between the outcome $Y$ (cUHDRS) and the censoring time $C$ that violates the outcome-independence assumption $C \independent Y \mid (X, \bZ)$ used by most existing estimators.

The inverse probability weighting (IPW) estimator has been used to address outcome-dependent censoring \citep{Lvetal2017, Atemetal2019}. The estimator up-weights uncensored observations based on their probability of being observed, but discards all censored observations in the final analytical sample. In Huntington disease studies where most participants exit before developing MCI, this exclusion of three-quarters of collected data (approximately 78\%) substantially reduces statistical power and limits precision for estimating disease trajectories.

In this paper, we develop three consistent estimators for outcome-dependent censoring that incorporate all observations: two augmented inverse probability weighting (AIPW) estimators and one maximum likelihood estimator (MLE). A key contribution is our guidance on estimator selection based on censoring rate---a design feature that varies substantially across studies and directly impacts estimator performance.  We show that no single estimator dominates. The MLE achieves the highest efficiency under low censoring rates (below 60\%), while the AIPW estimators perform best under moderate to high censoring (above 60\%). The censoring-rate-dependent performance means that the best estimator for one study may perform poorly in another---a practical reality not previously characterized. We establish consistency and asymptotic properties for all three estimators and, for one AIPW estimator, we establish a closed-form solution that eliminates the need for numerical integration, facilitating implementation. Applied to the PREDICT-HD (Neurobiological Predictors of Huntington Disease) study data with 78\% censoring, our framework demonstrates how censoring-rate-informed estimator selection improves precision in characterizing pre-MCI functional decline.

The remainder of the paper is organized as follows. Section~\ref{sec:chpt3-cautionary} presents a cautionary example showing how estimators designed for outcome-independent censoring produce bias under outcome-dependent censoring, motivating our methodological development. Section~\ref{sec:chp3-practical-estimators-dependent} introduces the three estimators and guides on selecting the preferred estimator based on censoring rate, establishing their consistency and asymptotic properties. Section~\ref{sec:chp3-simulations} presents simulation studies supporting our theoretical results. Section~\ref{sec:chp3-real_data} applies the proposed methods to PREDICT-HD data. Section~\ref{sec:chp3-discussion} concludes with broader implications and practical guidance for applied researchers.

\section{A cautionary note on existing estimators}
\label{sec:chpt3-cautionary}

The inverse probability weighting estimator was originally developed for outcome-independent censoring and successfully extended to outcome-dependent censoring \citep{Lvetal2017, Atemetal2019}. This success might suggest that other estimators designed for outcome-independent censoring could be similarly adapted. However, not all such estimators extend successfully. We demonstrate below that several estimators for outcome-independent censoring are biased under outcome-dependent censoring. To establish this result, we first introduce notation used throughout the paper. We use subscripts to distinguish among conditional densities and expectations. For example, $f_{Y|X,\bZ}(y, x, \bz; \btheta)$ denotes the conditional density of $Y$ given $(X, \bZ)$ evaluated at $(Y=y, X=x, \bZ=\bz)$, whereas $E_{Y|X,\bZ}(\cdot)$ denotes the conditional expectation with respect to $f_{Y|X,\bZ}$. Proofs of all propositions and theorems are provided in the \textit{Supplementary Material}.

\subsection{Biased estimators}
\label{sec:cautionary-note}

Two commonly used estimators for outcome-independent censoring are the complete case (CC) and the conditional mean imputation estimators. The CC estimator discards right-censored observations and solves the estimating equation
\bse
\sumi \delta_i \bS_{\btheta}^F (y_i,w_i,\bz_i; \btheta) = \bzero,
\ese
whereas the conditional mean imputation estimator replaces right-censored values of $X$ with their conditional expectations and solves the estimating equation
\bse
\sumi \delta_i \bS_{\btheta}^F (y_i,w_i,\bz_i; \btheta) + (1-\delta_i) \bS_{\btheta}^F \{y_i, E_{X|Y,\bZ, X>C}(X),\bz_i; \btheta\} = \bzero.
\ese
In both estimating equations, $\bS_\btheta^F(y,w,\bz;\btheta) = \partial \log f_{Y|X,\bZ}(y,w,\bz; \btheta) / \partial \btheta\trans$ denotes the score function of the outcome model. These estimators are straightforward to implement, but face challenges under outcome-dependent censoring. In our Huntington disease application, the CC estimator would discard the 78\% of participants who exit before MCI, while conditional mean imputation would replace their unobserved time-to-MCI with an estimated average. One might hope these estimators remain consistent despite the data loss or imputation, but Proposition~\ref{prop:chp3-cc-cmi-biased} shows they are biased under outcome-dependent censoring.

\begin{Prop}
\label{prop:chp3-cc-cmi-biased}
Assume $C \notindependent Y \mid (X, \bZ)$ and $C \independent X \mid (Y, \bZ)$. Then the complete case and conditional mean imputation estimators---where right-censored values are replaced by either $E_{X|Y, \bZ, X > C}(X)$ or $E_{X|\bZ}(X)$---are inconsistent estimators. Moreover, the augmented complete case and modified augmented complete case estimators are also inconsistent.
\end{Prop}
The inconsistency of the CC and conditional mean imputation estimators under outcome-dependent censoring has been previously established \citep{Bernhardtetal2014,Lvetal2017}. Proposition~\ref{prop:chp3-cc-cmi-biased} additionally shows that augmented variants of the CC estimator also fail under outcome-dependent censoring. These augmented estimators---including the augmented complete case (ACC) and modified augmented complete case (MACC) estimators \citep{vazquez2024establishing}---build upon the CC estimator by adding augmentation terms designed to improve efficiency under outcome-independent censoring. However, because these estimators retain the CC estimator as their base, they inherit its bias under outcome-dependent censoring.

To illustrate the consequences of these biases, consider what would occur if these estimators were mistakenly applied to our Huntington disease application. Estimating cUHDRS as a function of time to MCI onset using the CC, conditional mean imputation, ACC, or MACC estimators would yield biased results that distort the trajectory of decline. Individuals with higher baseline cUHDRS scores---indicating better health---experience slower cUHDRS decline and are more likely to be right-censored because they are less likely to reach MCI during follow-up. CC-based analyses would disproportionately exclude these slower-declining individuals, overrepresenting faster decliners and thereby overestimating the rate of decline. In contrast, conditional mean imputation, ACC, and MACC-based analyses attempt to incorporate censored observations using imputation or augmentation strategies that do not account for outcome-dependent censoring, leading to different but equally problematic biases. Such mischaracterization of the disease progression may lead to flawed assessments of treatment effects and erroneous trial design decisions.

\subsection{Unbiased but inefficient estimator}
\label{sec:chpt3-ipw}

The IPW estimator corrects the bias introduced by outcome-dependent censoring by assigning weights to uncensored observations based on the inverse of the probability of being observed. Formally, the IPW estimator solves the estimating equation
\bse
\sumi  \frac{\delta_i\bS_\btheta^F(y_i,w_i,\bz_i; \btheta)}{\pi_{Y,X,\bZ}(y_i,w_i,\bz_i)} =\bzero,
\ese 
where $\pi_{Y,X,\bZ}(y, w, \bz) =\pr(\Delta=1|Y=y,X=w,\bZ=\bz)= \int I(w<c) f_{C|Y,\bZ}(c,y,\bz) dc$. In the Huntington disease application, this weighting corrects for the fact that observations with higher cUHDRS scores are more likely to be right-censored, thus rebalancing the sample to represent the full population. Proposition~\ref{prop:chp3-ipw} establishes consistency.

\begin{Prop}
\label{prop:chp3-ipw}
Assume $C \notindependent Y \mid (X, \bZ)$ and $C \independent X \mid (Y, \bZ)$. The inverse probability weighting estimator is consistent under correct specification of $f_{C|Y,\bZ}$. 
\end{Prop}

While consistent, the IPW estimator's reliance on uncensored observations alone motivates the development of more efficient alternatives that incorporate all available data.

\section{Consistent and efficient estimators for outcome-dependent censoring}
\label{sec:chp3-practical-estimators-dependent} 

We develop three estimators that incorporate all observations: two AIPW estimators suited for high censoring rates and one maximum likelihood estimator for settings with low censoring.

\subsection{Augmented inverse probability estimators for high censoring rates}
\label{sec:chp3-estimators-high}

The AIPW estimator corrects the bias of the CC estimator under outcome-dependent censoring while improving upon the efficiency of the IPW estimator. AIPW estimators are widely used across missing data and causal inference problems to correct for bias and improve efficiency \citep{rotnitzky1997analysis, Tsiatis2006}. We introduce the form of the AIPW estimator tailored to the outcome-dependent censoring problem:
\bse
\sumi \bPhi_{\rm AIPW}(\bO_i; \btheta, \boldeta) = 
\sumi
\underbrace{
\frac{\delta_i\bS_\btheta^F(y_i,w_i,\bz_i; \btheta)}{\pi_{Y,X,\bZ}(y_i,w_i,\bz_i;\boldeta)}
}_{\text{IPW estimator}} 
+ 
\underbrace{
\left\{1- \frac{\delta_i}{\pi_{Y,X,\bZ}(y_i,w_i,\bz_i;\boldeta)}\right\} \bPsi_{\rm AIPW} (y_i,\bz_i; \btheta)
}_{\text{Augmentation Term}}  
= \bzero.
\ese
The first term is the IPW estimator, while the second is an augmentation term that incorporates information from censored observations. The augmentation component $\bPsi_{\rm AIPW}(y, \bz; \btheta)$ is a vector function of $(Y, \bZ)$ and $\btheta$. Under outcome-dependent censoring, the probability of observing $X$ depends on the outcome $Y$, which is why the weights use $\pi_{Y,X,\bZ}(y,w,\bz; \boldeta)$ rather than the simpler $\pr(\Delta=1|X=w,\bZ=\bz)$ used in outcome-independent settings \citep{vazquez2024establishing}. The AIPW estimator is consistent provided that $f_{C|Y,\bZ}$ (equivalently $\pi_{Y,X,\bZ}(y, w, \bz; \boldeta)$) is correctly specified (Theorem~\ref{thm:chp3-aipw}).

To implement the AIPW estimator, we first specify a parametric form for $f_{C|Y,\bZ}$ and estimate its parameters $\boldeta$. While any parametric model can be used, we adopt a Weibull accelerated failure time (AFT) model, which is commonly used for modeling time-to-event distributions in neurodegenerative disease studies and provides computational tractability \citep{swindell2009accelerated}. Specifically, we model $\log(C) = m_{\tilde\boldeta}(Y,\bZ;\tilde\boldeta) + \tilde\sigma_{\boldeta} \tilde\epsilon$, where $\tilde\epsilon \sim \text{Gumbel}(0,1)$, $m_{\tilde\boldeta}(\cdot)$ is a function of $(Y,\bZ)$ parametrized by $\tilde\boldeta$, and $\boldeta = (\tilde\boldeta\trans, \tilde\sigma_{\boldeta})\trans$. The maximum likelihood estimator of $\boldeta$ can be obtained using standard survival analysis software \citep{liu2023using} and solves
\be
 && \sumi \bPhi_\text{AFT}(\bO_i; \boldeta) \nonumber \\
  && = \sumi 
  \delta_i \frac{\int_{w_i\leq C} \bS^{F}_{\boldeta} (c,y_i,\bz_i; \boldeta) f_{C|Y,\bZ}(c,y_i,\bz_i; \boldeta) dc}{\int_{w_i\leq C} f_{C|Y,\bZ}(c,y_i,\bz_i; \boldeta) dc } + (1-\delta_i) \bS^{F}_{\boldeta} (w_i,y_i,\bz_i; \boldeta) = \bzero, \label{eqn:chp3-alpha_est}
\ee
where $\bS^{F}_{\boldeta} (c,y,\bz; \boldeta) = \partial \log f_{C|Y,\bZ} (c,y,\bz; \boldeta)/\partial \boldeta\trans$. The Weibull survival function then defines the probability of observing $X$ as
\bse
\pi_{Y,X,\bZ}(y,w,\bz; \boldeta) = \int_{w}^\infty f_{C|Y,\bZ}(c,y,\bz; \boldeta) dc = \exp\left( - \left[w/\exp\{m_{\tilde\boldeta}(y,\bz;\tilde\boldeta)\}\right]^{1/\tilde\sigma_{\boldeta}} \right).
\ese
Substituting the estimate $\wh\boldeta$ yields $\pi_{Y,X,\bZ}(y, w, \bz; \wh\boldeta)$ for use in the AIPW estimating equation. Theorem~\ref{thm:chp3-aipw} establishes the asymptotic distribution of the AIPW estimator when $\boldeta$ is estimated via \eqref{eqn:chp3-alpha_est}. The asymptotic distribution of the IPW estimator under the same weight estimation approach is also provided in the \textit{Supplementary Material}.

\begin{Th}
\label{thm:chp3-aipw}
Assume $C \notindependent Y \mid (X, \bZ)$ and $C \independent X \mid (Y, \bZ)$. Then, the augmented inverse probability weighting estimator is consistent and asymptotically normal, provided that $f_{C|Y,\bZ}$ is correctly specified. The asymptotic distribution of the AIPW estimator, when the probability of observing $X$ is indexed by nuisance parameter $\boldeta$ and estimated using \eqref{eqn:chp3-alpha_est},  is
\bse
n^{1/2}(\wh{\btheta}_\text{AIPW,AFT}-\btheta_0) \rightarrow_d \Normal(\bzero, \bA_\text{AIPW,AFT}^{-1}\bB_\text{AIPW,AFT}\bA_\text{AIPW,AFT}\trans), 
\ese
where 
\bse
&& \bA_\text{AIPW,AFT} = E_{Y,W,\Delta,\bZ} \left\{  \partial \bPhi_{\rm AIPW}(\bO; \btheta,\boldeta_0)/ \partial \btheta\trans |_{\btheta = \btheta_0}
\right\}; \\
&& \bB_\text{AIPW,AFT} =  E\{(\bPhi_{\rm AIPW}(\bO; \btheta_0,\boldeta_0) \\
&& \quad - E\{\partial\bPhi_{\rm AIPW}(\bO;\btheta_0,\boldeta)/\partial\boldeta^T |_{\boldeta = \boldeta_0} \}[E\{\partial\bPhi_\text{AFT}(\bO;\boldeta)/\partial\boldeta^T |_{\boldeta = \boldeta_0}\}]^{-1}\bPhi_\text{AFT}(\bO;\boldeta_0) )^{\otimes2}\}.
\ese
An estimate for the variance of $\wh{\btheta}_\text{AIPW,AFT}$ is obtained by replacing $(\btheta_0\trans, \boldeta_0\trans)$ with their estimated counterparts $(\wh{\btheta}_\text{AIPW,AFT}\trans, \wh\boldeta\trans)$ in $\bA_{\rm AIPW;\boldeta}^{-1}\bB_{\rm AIPW;\boldeta}\bA_{\rm AIPW;\boldeta}\trans$. 
\end{Th}

The AIPW estimator's consistency requires only correct specification of $f_{C|Y,\bZ}$. Under high censoring rates, this requirement is practical because the triplet $(Y,C,\bZ)$ is observed for most participants. For example, in PREDICT-HD where 78\% of participants exit before MCI, censoring times are observed for the majority while time to MCI is observed for only 22\%. This pattern makes $f_{C|Y,\bZ}$ easier to estimate reliably than $f_{X|Y,\bZ}$, which depends on the rarely-observed triplet $(Y,X,\bZ)$. Therefore, under high censoring, the AIPW estimator achieves consistency by leveraging the frequently observed data.

Efficiency gains over the IPW estimator depend on the choice of augmentation term $\bPsi_{\rm AIPW}(y,\bz;\btheta)$. Under mild regularity conditions, any choice of $\bPsi_{\rm AIPW}(y,\bz;\btheta)$ yields consistency provided that $f_{C|Y,\bZ}$ is correctly specified, but not all choices improve efficiency. In the \textit{Supplementary Material}, we show that
\be
\label{eq:chp3-psi-aipw-eff}
\bPsi^{\rm eff}_\text{AIPW}(y, \bz; \btheta) = \frac{E_{X|Y,\bZ}\left[{1-1/\pi_{Y,X,\bZ}(y,X,\bz;\boldeta)}, \bS^F_{\btheta}(y,X,\bz;\btheta)\right]}{E_{X|Y,\bZ}\left[1-1/\pi_{Y,X,\bZ}(y,X,\bz;\boldeta)\right]},
\ee
guarantees efficiency gains relative to the IPW estimator when correctly specified. The trade-off, however, is that in addition to requiring correct specification of $f_{C|Y,\bZ}$ to compute $\pi_{Y,X,\bZ}(y,w,\bz;\boldeta)$, one must also correctly specify the conditional density $f_{X|Y,\bZ}$, which is needed to evaluate the conditional expectations in \eqref{eq:chp3-psi-aipw-eff}. Under high censoring, $f_{X|Y,\bZ}$ is difficult to estimate reliably because $(Y,X,\bZ)$ is rarely observed.

To address this limitation, we develop an updated augmentation that improves efficiency over the IPW estimator even when $f_{X|Y,\bZ}$ is misspecified. The updated form takes the form
$\bPsi_\text{updated}(y,\bz;\btheta) = \bLambda_\text{AIPW} \bPsi_\text{AIPW}(y,\bz;\btheta)$,
where $\bLambda_\text{AIPW}$ is a non-random full-rank matrix whose form is chosen to improve efficiency over the IPW estimator for any choice of $\bPsi_\text{AIPW}(y,\bz;\btheta)$ and \eqref{eqn:chp3-alpha_est}. We construct $\bLambda_\text{AIPW}$ as
\bse
\bLambda_{\rm AIPW} &=& -\biggr\{ E_{Y,W,\Delta,\bZ} \biggr( \left[ \left\{1-\Delta/\pi_{Y,X,\bZ}(Y,W,\bZ;\boldeta) \right\} \bPsi_{\rm AIPW}(Y,\bZ; \btheta) + \bA^*_\text{AIPW,AFT} \bUpsilon_\text{AFT}(\bO) \right] \\
&& \sextant \times \left\{ \Delta \bS_{\btheta}^F(Y,W,\bZ;\btheta)/\pi_{Y,X,\bZ}(Y,W, \bZ; \boldeta) + \bA^*_\text{IPW,AFT} \bUpsilon_\text{AFT}(\bO) \right\} \biggr) \biggr\}^T \\
&& \times E_{Y,W,\Delta,\bZ} \biggr( \biggr[ \left\{1- \Delta/\pi_{Y,X,\bZ}(Y,W, \bZ; \boldeta) \right\} \bPsi_{\rm AIPW}(Y,\bZ; \btheta)  + \bA^*_\text{AIPW,AFT} \bUpsilon_\text{AFT}(\bO) \biggr]^{\otimes 2} \biggr) ^{-T},
\ese 
where 
\bse
\bA^*_\text{IPW,AFT} &=& E_{Y,W,\Delta,\bZ} \left\{   \frac{\partial}{\partial \boldeta^T}  \frac{\Delta \bS^F_\btheta(Y,W,\bZ;\btheta)}{\pi_{Y,X,\bZ}(Y,W, \bZ; \boldeta)} \right\},\\
\bA^*_\text{AIPW,AFT} &=& E_{Y,W,\Delta,\bZ} \left( \frac{\partial}{\partial \boldeta^T} \biggr[\biggr\{1- \frac{\Delta}{\pi_{Y,X,\bZ}(Y,W, \bZ; \boldeta)} \biggr\} \bPsi_{\rm AIPW}(Y, \bZ; \btheta) \biggr] \right), \text{\ and} \\
\bUpsilon_\text{AFT}(\bO) &=& E_{Y,W,\Delta,\bZ} \left\{ \frac{\partial}{\partial\boldeta^T} \bPhi_\text{AFT}(\bO; \boldeta) \right\}^{-1} \bPhi_\text{AFT}(\bO; \boldeta).
\ese 
An estimate for $\bLambda_\text{AIPW}$ can be obtained by substituting consistent estimates of $(\btheta\trans, \boldeta\trans)$. Because $\bLambda_\text{AIPW}$---and by extension $\bPsi_\text{updated}(y,\bz;\btheta)$---depends on the choice of $\bPsi_\text{AIPW}(y,\bz; \btheta)$, any form satisfying mild regularity conditions yields efficiency gains over the IPW estimator.

For practical implementation, we recommend using a simpler choice for $\bPsi_\text{AIPW}(y,\bz;\btheta)$ within the $\bPsi_\text{updated}(y,\bz;\btheta)$ framework: the special case $\bPsi_\text{closed}(y,\bz; \btheta) = E_{X|Y,\bZ}\{\bS_{\btheta}^F (y,X,\bz; \btheta)\}$, which offers two advantages. First, it requires evaluating only one conditional expectation rather than two compared to \eqref{eq:chp3-psi-aipw-eff}. Second, when $X$ follows a normal distribution and $m(X,\bZ; \btheta)$ is linear in $X$, this term has a closed-form solution that eliminates numerical integration entirely \citep{vazquez2024establishing}. This closed form is particularly valuable in applications like Huntington disease research where analysts may fit multiple models to explore different covariate specifications. When $m(\cdot)$ is nonlinear in $X$, $\bPsi_\text{closed}$ may not have a closed-form expression but remains computationally simpler than \eqref{eq:chp3-psi-aipw-eff}. Importantly, the consistency of the AIPW estimator with $\bPsi_\text{closed}(y,\bz;\btheta)$ depends only on correct specification of $f_{C|Y,\bZ}$, not $f_{X|Y,\bZ}$, so misspecification of the normality assumption for $X$ does not affect consistency (Theorem~\ref{thm:chp3-aipw}).

Table~\ref{tab:efficiency-augmented} summarizes the density requirements for each augmentation term. The efficient augmentation $\bPsi_\text{AIPW}^\text{eff}(y,\bz;\btheta)$ requires correct specification of both $f_{C|Y,\bZ}$ and $f_{X|Y,\bZ}$, while our recommended approach using $\bPsi_\text{closed}(y,\bz;\btheta)$ within $\bPsi_\text{updated}(y,\bz;\btheta) = \bLambda_\text{AIPW} \bPsi_\text{closed}(y,\bz;\btheta)$ achieves efficiency gains without requiring correct specification of $f_{X|Y,\bZ}$ nor $f_{C|Y,\bZ}$ when constructing the augmentation term. Note that correct specification of $f_{C|Y,\bZ}$ remains necessary for the overall consistency of the AIPW estimator, regardless of which augmentation term is used.

\begin{table}[ht]
\centering
\begin{tabular}{lccc}
\toprule
 & $f_{C|Y,\bZ}$ & $f_{X|Y,\bZ}$ & $f_{Y|X,\bZ}$ \\
\midrule
$\bPsi_\text{AIPW}^\text{eff}(y, \bz;\btheta)$ & $\times$ & $\times$ & $\times$ \\
$\bPsi_\text{updated}(y, \bz;\btheta)$ &   --   &   --   & $\times$ \\
\bottomrule
\end{tabular}
\caption{\textbf{Required densities for correct specification of the augmentation term.} A ``$\times$'' indicates densities required for correct specification of the augmentation term to achieve efficiency gains relative to the IPW estimator. Although $f_{C|Y,\bZ}$ is not required for specifying the augmentation term in $\bPsi_\text{updated}(y,\bz;\btheta)$, it remains necessary for the overall consistency of the AIPW estimator.\label{tab:efficiency-augmented}}
\end{table}

\subsection{Maximum likelihood estimator for low censoring rates}
\label{sec:chp3-estimators-low}

Under low censoring rates, most participants reach the event (e.g., MCI onset in Huntington disease) before study exit, so the triplet $(Y,X,\bZ)$ is observed more frequently than $(Y,C,\bZ)$. With event times frequently observed, $f_{X|\bZ}$ can be estimated reliably, whereas $f_{C|Y,\bZ}$ becomes difficult to estimate. The MLE requires correct specification of $f_{X|\bZ}$ for consistency, making it better-suited for low censoring settings.

To implement the MLE, we specify $f_{X|\bZ}$ using a Weibull AFT model and estimate its finite parameter vector $\bgamma$ from the observed $(W,\Delta,\bZ)$ data. The approach parallels the AIPW estimator's use of a Weibull AFT model for $f_{C|Y,\bZ}$, but with the roles of $f_{X|\bZ}$ and $f_{C|Y,\bZ}$ reversed. The MLE solves the estimating equation
\bse
 \sumi \bPhi_{\rm MLE}(\bO_i; \btheta, \bgamma) = 
 \sumi \underbrace{\delta_i \bS_{\btheta}^F (y_i,w_i,\bz_i; \btheta)}_\text{CC Estimator}  + \underbrace{(1-\delta_i) \frac{\int_{I(w_i < X)} \bS_{\btheta}^F (y_i,x,\bz_i; \btheta) f_{Y\mid X,\bZ}(y_i,x,\bz_i; \btheta) f_{X \mid \bZ}(x, \bz_i; \bgamma) dx}{\int_{I(w_i < X) } f_{Y\mid X,\bZ}(y_i,x,\bz_i; \btheta) f_{X \mid \bZ}(x,\bz_i;\bgamma) dx}}_\text{Augmentation Term} = \bzero,
\ese 
where the first term is the CC estimator and the second is an augmentation term that incorporates information from censored observations. Unlike the ACC and MACC estimators which also augment the CC estimator but fail to correct its bias under outcome-dependent censoring, the MLE's augmentation term ensures consistency provided that $f_{X|\bZ}$ is correctly specified. Theorem~\ref{thm:chp3-mle} establishes asymptotic normality when $\bgamma$ is estimated using a Weibull AFT model analogous to \eqref{eqn:chp3-alpha_est}.

\vspace{1em}
\begin{Th}
\label{thm:chp3-mle}
Assume $C \notindependent Y \mid (X,\bZ)$ and $C \independent X \mid (Y,\bZ)$. 
Then, the MLE is consistent and asymptotically normal, provided that $f_{X|\bZ}$ is correctly specified. The asymptotic distribution, when $f_{X|\bZ}$ is indexed by $\bgamma$ and estimated using an AFT model of the form \eqref{eqn:chp3-alpha_est}, is
\bse
n^{1/2}(\wh{\btheta}_\text{MLE,AFT} - \btheta_0) \to_d 
\Normal\!\left(\bzero, \bA_\text{MLE,AFT}^{-1}\,\bB_\text{MLE,AFT}\,\bA_\text{MLE,AFT}\trans\right),
\ese
where 
\bse
&& \bA_\text{MLE,AFT} = E\left\{  \partial \bPhi_{\rm MLE}(\bO; \btheta,\bgamma_0)/ \partial \btheta\trans |_{\btheta = \btheta_0}
\right\}; \\
&& \bB_\text{MLE,AFT} =  E\{(\bPhi_{\rm MLE}(\bO; \btheta_0,\bgamma_0) \\
&& \quad - E\{\partial\bPhi_{\rm MLE}(\bO;\btheta_0,\bgamma)/\partial\bgamma^T |_{\bgamma = \bgamma_0} \}[E\{\partial\bPhi_\text{AFT}(\bO;\bgamma)/\partial\bgamma^T |_{\bgamma = \bgamma_0}\}]^{-1}\bPhi_\text{AFT}(\bO;\bgamma_0) )^{\otimes2}\}.
\ese
An estimate for the variance of $\wh{\btheta}_\text{MLE,AFT}$ is obtained by replacing $(\btheta_0\trans,\bgamma_0\trans)$ with their estimated counterparts 
$(\wh{\btheta}_\text{MLE,AFT}\trans,\wh\bgamma\trans)$ in 
$\bA_\text{MLE,AFT}^{-1}\,\bB_\text{MLE,AFT}\,\bA_\text{MLE,AFT}\trans$.
\end{Th}

Theorems~\ref{thm:chp3-aipw} and \ref{thm:chp3-mle} formalize the intuition that estimator performance depends on censoring rate through data availability. Under high censoring, $(Y,C,\bZ)$ is frequently observed, favoring AIPW estimators that require $f_{C|Y,\bZ}$. Under low censoring, $(Y,X,\bZ)$ is frequently observed, favoring the MLE that requires $f_{X|\bZ}$. Corollary~\ref{cor:low-high-censoring} summarizes these recommendations, which are further validated through simulation studies in Section~\ref{sec:chp3-simulations}.

\vspace{1em}
\begin{Coro}
\label{cor:low-high-censoring}
\textit{Assume $C \notindependent Y \mid (X,\bZ)$ and $C \independent X \mid (Y,\bZ)$. (i) Under high censoring rates, the IPW and AIPW estimators are recommended, as their consistency requires correct specification of $f_{C|Y,\bZ}$. (ii) Under low censoring rates, the MLE is recommended, as its consistency requires correct specification of $f_{X|\bZ}$.}
\end{Coro}

\section{Simulation Study}
\label{sec:chp3-simulations}

To support our theoretical results, we conducted a simulation study to assess the robustness of the estimators, focusing on their consistency and efficiency under low to high censoring rates. 

\subsection{Data generation and metrics of comparison}

We considered the regression model
\bse
Y = \beta_{0} + \beta_{AX} (A - X) + \beta_{Z} Z + \epsilon, \quad \epsilon \sim \Normal(0, \sigma^2),
\ese
where $A$ represents the current age, $X$ is the age at MCI onset, and $A - X$ represents the time to MCI onset. We simulated $A \sim \Normal(2, 1)$ and $Z \sim \Normal(0, 1)$. Then, $X$ was drawn from a Weibull distribution with scale parameter $\exp(\tilde\gamma_0 + \tilde\gamma_{Z} Z)$ and shape parameter $1/\tilde\sigma_\gamma$, such that $(\tilde\gamma_0, \tilde\gamma_Z, \tilde\sigma_\gamma)^T = (0.1, 0.1, 0.5)^T$. The outcome $Y$ was then computed using the parameter set $\btheta = (\beta_0, \beta_{AX}, \beta_{Z}, \sigma)^T = (1, 1, 1, 1)^T$. Finally, the censoring time $C$ was drawn from a Weibull distribution with scale parameter $\exp(\tilde\eta_0 + \tilde\eta_{Y} Y + \tilde\eta_{Z} Z)$ and shape parameter $1/\tilde\sigma_\eta$, using $(\tilde\eta_0, \tilde\eta_Y, \tilde\eta_Z, \tilde\sigma_\eta)^T = (\tilde\eta_0, 0.5, 0.5, 1.5)^T$, where $\tilde\eta_0$ was varied to achieve censoring rates between 10\% and 95\%. A total of $N = 3{,}000$ simulations with $n = 1{,}000$ observations were performed. 

We evaluated the IPW estimator, both AIPW variants (one using $\bPsi^{\text{eff}}_{\text{AIPW}}$ and the other using $\bPsi_{\text{updated}}$ with the robustness matrix $\bLambda$), and the MLE under varying censoring rates (10\%--95\%). To demonstrate the bias introduced by inconsistent estimators (Proposition~\ref{prop:chp3-cc-cmi-biased}), we also evaluated the CC, conditional mean imputation (replacing unobserved $X$ with either $E_{X|\bZ}(X)$ or $E_{X|Y,\bZ,\Delta=0}(X)$), ACC, MACC, and naive estimators (treating $W$ as $X$). We included the oracle estimator as a benchmark representing no censoring.

At 60\% censoring, we examined three specification scenarios for the consistent estimators: (i) correct (nuisance parameters set to true values), (ii) estimated (nuisance parameters estimated from data using Weibull AFT models), and (iii) incorrect (misspecified nuisance parameters). For IPW and AIPW, incorrect specification involved random weights from $1/\text{uniform}(0.1, 0.9)$ rather than correctly specifying $f_{C|Y,\bZ}$ (equivalently, $\boldeta$). For AIPW using $\bPsi^{\text{eff}}(y,\bz;\btheta)$, we additionally considered incorrect specification of the augmentation term as $E_{X|Y,\bZ}\{\bS_{\btheta}^F(y,X,\bz;\btheta)\}$. For MLE, incorrect specification involved shape parameter $\tilde\sigma_\gamma = 1.2$ (misspecifying $f_{X|\bZ}$, equivalently $\bgamma$).

For the $k^{th}$ element of $\btheta$, we computed: percent bias ${[\bar\theta^{k} - \theta^{k}_0]/\theta^{k}_0} \times 100$, where $\bar\theta^{k} = \sum_{i=1}^{N} \wh\theta^{k}_i / N$; mean estimated standard errors ${\rm SE} = \sum_{i=1}^{N} \wh{\rm SE}(\theta^k)_i / N$; empirical standard errors \newline $\wh{\rm SE} = \sqrt{\sum_{i=1}^{N} (\wh\theta^{k}_i - \bar\theta^{k})^2 / N}$; and nominal coverage ${\rm Cov} = \sum_{i=1}^{N} I(\theta^{k}_0 \in \text{Wald 95\% CI}) / N$.

\subsection{Simulation results}

Our simulations demonstrated that no single estimator dominates across all censoring rates, corroborating Corollary~\ref{cor:low-high-censoring}. Figure~\ref{fig:nuisance} shows that estimator performance depends on censoring rate. Under low censoring (below approximately 60\%), all estimators achieved bias near 0\% and coverage close to 95\%. As censoring increased beyond 60\%, the MLE's performance deteriorated substantially while IPW and AIPW estimators maintained lower bias and higher nominal coverage. For example, at 95\% censoring, the percent bias for IPW and AIPW remained within about 5\% (for $\beta_0$), whereas the MLE exceeded 20\%, resulting in coverage probabilities below 50\%. This pattern reflects that under high censoring, $(Y,C,\bZ)$ is frequently observed making $f_{C|Y,\bZ}$ estimable, whereas $(Y,X,\bZ)$ is rarely observed making $f_{X|\bZ}$ difficult to estimate reliably. The IPW and AIPW estimators performed well even under low censoring despite $f_{C|Y,\bZ}$ being harder to estimate in this setting, because the small proportion of censored observations minimizes the impact of any misspecification.

{
\spacingset{1}
\begin{figure}[!h]
    \centering
    \includegraphics[width=0.9\textwidth]{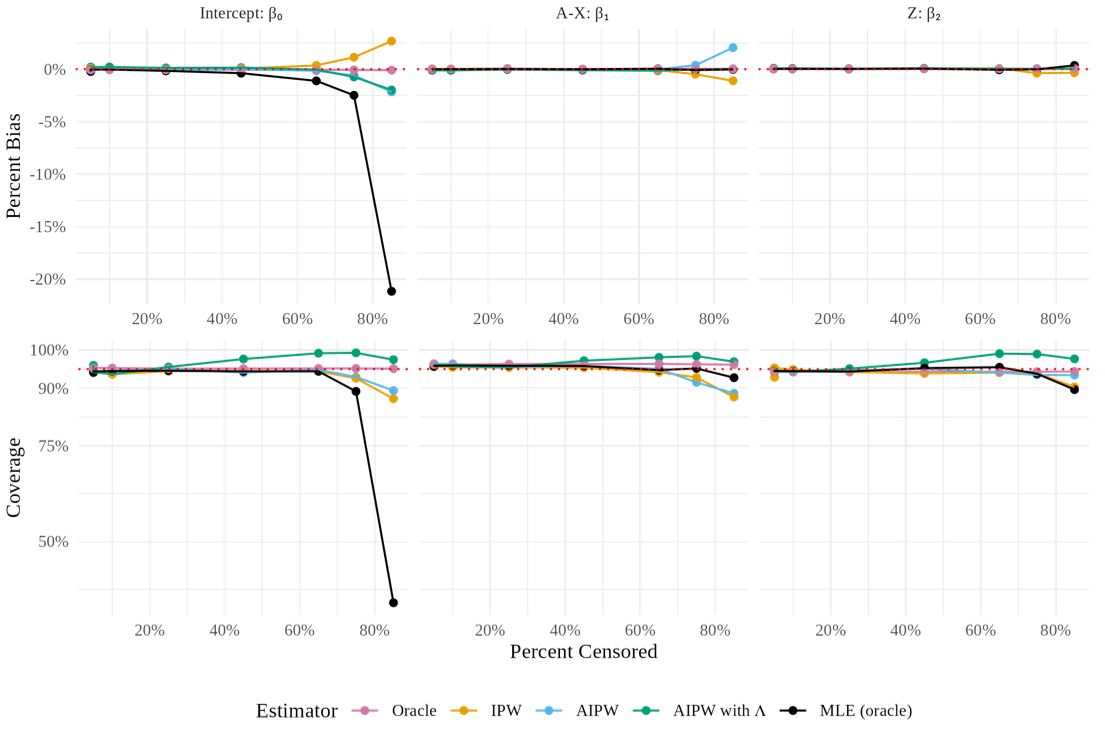}
    \caption{\textbf{Bias and coverage by censoring rate.} The red dashed line represents 0\% bias in the first panel and 95\% nominal coverage in the second panel.}
    \label{fig:nuisance}
\end{figure}
}

To examine estimator behavior at moderate-to-high censoring we conducted additional simulations at 60\% censoring under the three specification scenarios. When nuisance parameters were correctly specified, all consistent estimators achieved bias close to 0\% and coverage near 95\%, supporting the theoretical claims of Propositions~\ref{prop:chp3-cc-cmi-biased}--\ref{prop:chp3-ipw} and Theorems~\ref{thm:chp3-aipw}--\ref{thm:chp3-mle} (Table~\ref{tab:chp3-sims_1000}). Misspecification revealed the importance of correct model specification: when weights were incorrectly specified, IPW and AIPW exhibited high bias and poor coverage; similarly, MLE performed poorly when $f_{X|\bZ}$ was misspecified.

Efficiency comparisons at 60\% censoring showed that AIPW and MLE can improve upon IPW when correctly specified. For the AIPW estimator without $\bLambda$, efficiency gains over IPW required correct specification of both the weights and $\bPsi(y,\bz; \btheta)$; when only weights were correct, AIPW remained approximately unbiased but was less efficient than IPW. The AIPW estimator with $\bLambda$ maintained higher efficiency than IPW when weights were correctly specified, regardless of $\bPsi$ specification, while also being computationally faster than the version using $\bPsi^{\text{eff}}$. When $f_{X|\bZ}$ was correctly specified, MLE achieved the highest efficiency among all estimators, though at 60\% censoring this advantage was modest given the difficulty of estimating $f_{X|\bZ}$ reliably with relatively few uncensored observations.

\begin{table}[!h]
\centering
\resizebox{1\linewidth}{!}{
\begin{tabular}[t]{llrrrrrrrrrrrrrr}
\toprule
\textbf{Estimator} & \textbf{Specification} &  \textbf{Bias} & \textbf{SE} & \textbf{SD} & \textbf{Cov} & \textbf{Bias} & \textbf{SE} & \textbf{SD} & \textbf{Cov} &  \textbf{Bias} & \textbf{SE} & \textbf{SD} & \textbf{Cov} \\
\midrule
\addlinespace[0.3em]
&&\multicolumn{4}{c}{\textbf{Intercept: $\beta_0 = 1$}} &\multicolumn{4}{c}{\textbf{$A-X$: $\beta_{AX} = 1$}} &\multicolumn{4}{c}{\textbf{$Z$: $\beta_{Z} = 1$}}\\
\addlinespace
\textbf{Oracle} & &  -0.09 & 3.25 & 3.30 & 94.31 & -0.01 & 1.38 & 1.38 & 95.02 & 0.01 & 3.18 & 3.17 & 94.68 \\
\addlinespace
\multicolumn{14}{l}{\textbf{Consistent Estimators}} \\
\addlinespace
 & correct $\pi_{Y,X,\bZ}$ & -0.21 & 6.13 & 6.37 & 93.61 & -0.10 & 2.25 & 2.37 & 94.05 & -0.10 & 5.70 & 5.87 & 94.62 \\
\multirow[t]{-2}{*}{\raggedright\arraybackslash \hspace{1em}IPW} 
& incorrect $\pi_{Y,X,\bZ}$ & -21.89 & 6.17 & 6.30 & 6.19 & -1.26 & 2.25 & 2.37 & 89.33 & -0.30 & 5.82 & 5.96 & 94.55 \\
& estimated  $\pi_{Y,X,\bZ}$ & -0.07 & 5.61 & 5.62 & 94.05 & -0.13 & 2.19 & 2.27 & 93.84 & -0.07 & 5.61 & 5.96 & 94.15 \\
\addlinespace
 & correct $f_{X|\bZ}$& -0.06 & 4.03 & 3.99 & 95.41 & -0.04 & 1.72 & 1.70 & 94.31 & 0.02 & 4.00 & 4.04 & 94.78 \\
\multirow[t]{-2}{*}{\raggedright\arraybackslash \hspace{1em}MLE} 
& incorrect $f_{X|\bZ}$  & -5.86 & 4.01 & 3.97 & 68.45 & -3.25 & 1.76 & 1.77 & 55.22 & -1.54 & 3.99 & 4.03 & 92.57 \\
& estimated $f_{X|\bZ}$  & 0.10 & 4.18 & 4.01 & 95.82 & -0.01 & 1.69 & 1.74 & 93.98 & 0.07 & 4.09 & 4.17 & 94.29 \\
\addlinespace
&\multicolumn{5}{l}{\underline{without using $\bLambda$}}\\
\multirow[t]{-2}{*}{\raggedright\arraybackslash \hspace{1em}AIPW} 
 & correct $\pi_{Y,X,\bZ}$ and correct $\bPsi_{\rm AIPW}(y,\bz)$  &  0.11 & 5.15 & 5.14 & 94.52 & 0.08 & 1.94 & 2.00 & 93.98 & 0.09 & 5.13 & 5.13 & 94.88 \\
& incorrect $\pi_{Y,X,\bZ}$ and correct $\bPsi_{\rm AIPW}(y,\bz)$  & -13.45 & 6.53 & 6.50 & 46.29 & -0.42 & 2.27 & 2.40 & 92.31 & 1.41 & 6.63 & 6.84 & 93.48 \\ 
 & correct $\pi_{Y,X,\bZ}$ and incorrect $\bPsi_{\rm AIPW}(y,\bz)$   & -0.17 & 11.49 & 11.40 & 94.85 & 0.01 & 4.09 & 4.08 & 94.85 & -0.13 & 10.08 & 10.26 & 94.62 \\ 
& incorrect $\pi_{Y,X,\bZ}$ and incorrect $\bPsi_{\rm AIPW}(y,\bz)$  & 25.50 & 15.29 & 15.66 & 63.41 & 0.10 & 3.53 & 3.78 & 92.98 & 11.43 & 14.26 & 14.72 & 89.36 \\ 
& estimated  $\pi_{Y,X,\bZ}$ and estimated $\bPsi_{\rm AIPW}(y,\bz)$ & 0.05 & 4.75 & 4.67 & 94.65 & 0.05 & 1.85 & 1.95 & 93.64 & 0.02 & 4.73 & 5.12 & 92.03 \\
&\multicolumn{5}{l}{\underline{using $\bLambda$}}\\
 & correct $\pi_{Y,X,\bZ}$  & -0.63 & 4.99 & 5.11 & 93.95 & -0.19 & 1.93 & 2.00 & 93.68 & -0.09 & 4.85 & 5.01 & 94.18 \\ 
& incorrect $\pi_{Y,X,\bZ}$  & -16.84 & 5.34 & 5.48 & 12.61 & -1.05 & 2.03 & 2.14 & 90.10 & 0.24 & 5.45 & 5.66 & 93.78 \\ 
& estimated $\pi_{Y,X,\bZ}$  & -0.57 & 5.47 & 4.79 & 96.27 & -0.22 & 2.09 & 1.96 & 95.06 & -0.13 & 5.38 & 5.04 & 95.46 \\
\addlinespace
\multicolumn{14}{l}{\textbf{Inconsistent Estimators}} \\
\addlinespace
\hspace{1em} Naive &   &  -98.90 & 5.00 & 4.97 & 0.00 & -18.25 & 1.97 & 1.97 & 0.00 & -10.79 & 5.25 & 5.31 & 47.22 \\ 
\addlinespace
\hspace{1em} CC & &   -21.78 & 5.13 & 5.25 & 1.37 & -1.26 & 1.90 & 1.98 & 88.19 & -0.17 & 4.87 & 4.94 & 94.55 \\ 
\addlinespace
\hspace{1em} ACC & using $\bLambda$ and correct  $\pi_{Y,X,\bZ}$ &   -19.13 & 4.61 & 5.12 & 2.81 & -1.31 & 1.84 & 1.90 & 87.93 & -8.20 & 5.01 & 10.37 & 52.78 \\ 
\addlinespace
\hspace{1em} MACC & using $\bLambda$ and correct  $\pi_{Y,X,\bZ}$ &  -21.79 & 4.61 & 4.66 & 0.40 & -1.25 & 1.83 & 1.90 & 88.26 & -0.46 & 6.06 & 6.28 & 95.92 \\ 
\addlinespace
\hspace{1em} MCI & correct $E_{X|Y,\bZ, \Delta=0}(X)$ & 2.75 & 2.81 & 7.73 & 63.31 & 4.92 & 1.68 & 14.33 & 0.64 & 1.27 & 2.65 & 4.61 & 86.19 \\ 
& correct $E_{X|\bZ}(X)$ & 36.37 & 6.00 & 5.90 & 0.00 & -2.03 & 2.41 & 2.45 & 86.86 & 6.07 & 5.67 & 5.73 & 80.94 \\ 
\addlinespace
\bottomrule
\end{tabular}}
\caption{\textbf{Simulation results of regression analysis with a right-censored covariate, 60\% right-censoring.} Simulation results of the mean estimate, mean estimated standard errors (SE), and empirical standard deviations (SD) scaled by 100, and 95\%  confidence interval coverage (Cov) for each estimator.  \label{tab:chp3-sims_1000}}
\end{table}

\begin{figure}[!h]
    \centering
    \includegraphics[width=0.85\textwidth]{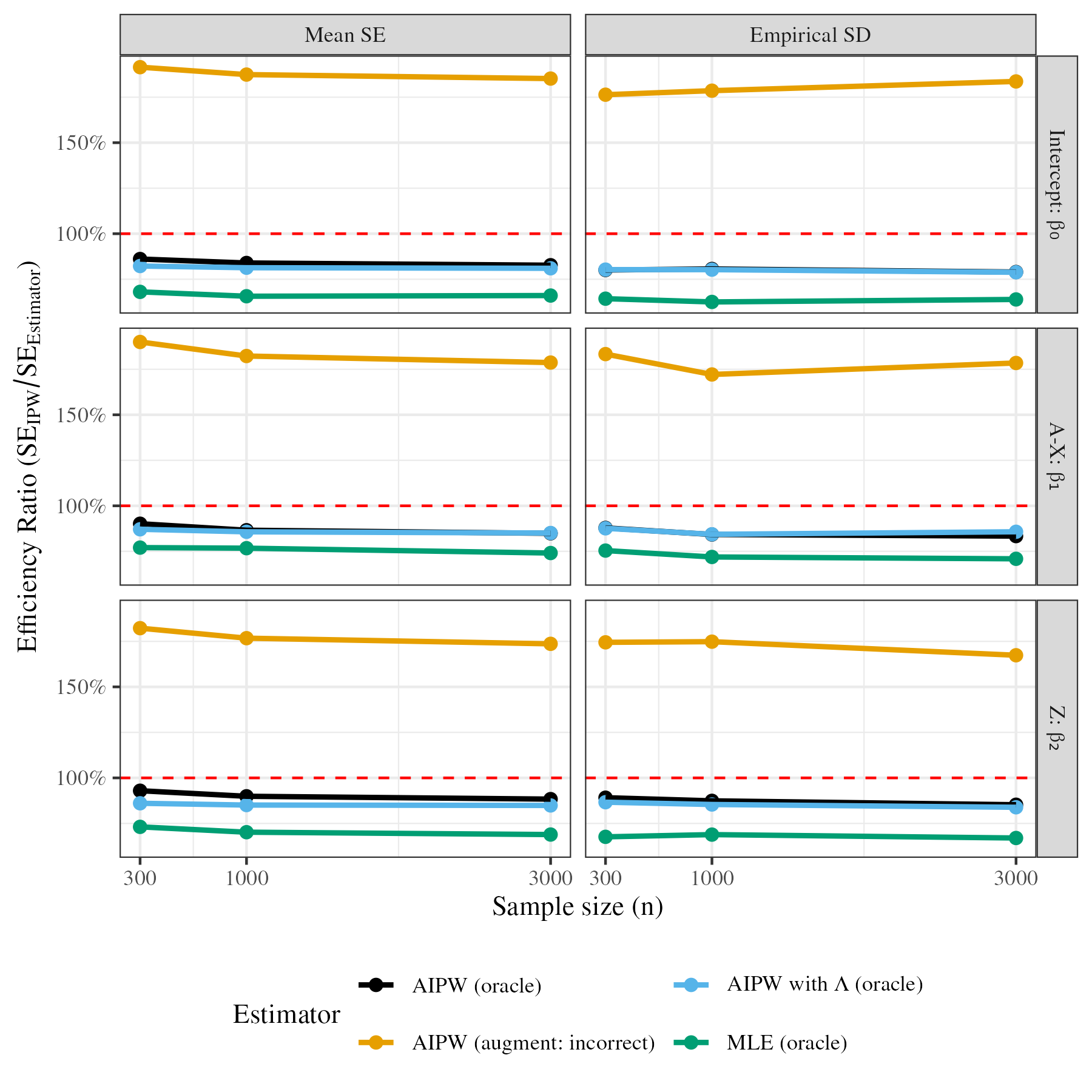}
    \caption{\textbf{Efficiency comparison of estimators against the IPW estimator.} The red-dashed line represents no gain in efficiency compared to the IPW estimator. Values below this line ($<100$\%) indicate greater efficiency, while values above it ($>100$\%) indicate reduced efficiency.  The empirical mean of the estimated standard errors (Mean SE) and the empirical standard deviation (Empirical SD) of $\wh\btheta$ across all simulations are illustrated.}
    \label{fig:efficiency}
\end{figure}

The efficiency patterns held consistently across sample sizes. Figure~\ref{fig:efficiency} illustrates the relative efficiency of the AIPW estimator (with and without $\bLambda$) and MLE compared to IPW at 60\% censoring for sample sizes 300, 1000, and 3000. The red-dashed line represents 100\% (no efficiency gain relative to IPW); values below indicate greater efficiency. Both AIPW variants and MLE exhibited higher efficiency than the IPW estimator across all sample sizes when nuisance parameters were correctly specified.

Finally, our simulations showed that commonly used alternatives fail under outcome-dependent censoring. The naive, CC, conditional mean imputation, ACC, and MACC estimators all exhibited substantial bias (Table~\ref{tab:chp3-sims_1000}), with the naive estimator showing bias exceeding 98\% for $\beta_0$ and the CC estimator exceeding 21\%. These results demonstrate the importance of accounting for outcome-dependent censoring when selecting estimation methods.

\section{Application to Huntington Disease}
\label{sec:chp3-real_data}

\subsection{Data and model specification}

We applied our estimators to PREDICT-HD to model the composite Unified Huntington Disease Rating Scale (cUHDRS) as a function of time to mild cognitive impairment (MCI) onset. The cUHDRS integrates cognitive, motor, and functional assessments and serves as a primary outcome in Huntington disease clinical trials \citep{rodrigues2019huntington}. Huntington disease is a genetic neurodegenerative disorder caused by cytosine-adenine-guanine (CAG) repeat expansions in the huntingtin gene, where individuals with CAG repeat length $\geq 40$ are guaranteed to develop the disease. Our analytical sample included 833 PREDICT-HD participants with CAG $\geq 40$, no MCI at baseline, and complete data. Of these participants, 184 developed MCI during follow-up, yielding 78\% censoring.

We modeled cUHDRS as
\bse
{\rm cUHDRS_i} = \beta_0 + \beta_{\rm TTMCI} {\rm TTMCI_i} +  \beta_{\rm CAG} {\rm CAG_i} + \beta_{\rm Education} {\rm Education_i} + \beta_{\rm Sex} {\rm I(Sex_i = Female)} + \epsilon_i,
\ese
adjusting for factors known to influence disease progression. Here, ${\rm TTMCI_i}$ is time to MCI (current age minus age at MCI onset); ${\rm CAG_i}$ is CAG repeat length centered at 40, where longer repeats confer greater disease burden and faster progression; ${\rm Education_i}$ is years of education centered at 14.8, where higher education is associated with cognitive reserve that may delay decline \citep{stern2002cognitive}; and ${\rm I(Sex_i = Female)}$ indicates biological sex. The cUHDRS outcome was centered at its mean of 17.5.

Censoring in PREDICT-HD is likely to be outcome-dependent (violating $C \independent Y \mid (X,\bZ)$), as cUHDRS scores influence both MCI diagnosis and dropout. At 78\% censoring, Corollary~\ref{cor:low-high-censoring} suggests the AIPW estimators should outperform MLE. We estimated parameters using IPW, both AIPW variants, and MLE, with $f_{C|Y,\bZ}$ and $f_{X|\bZ}$ estimated via Weibull AFT models (see \textit{supplementary material} for diagnostics).

\subsection{Results}
\label{sec:chp3-real_data_results}

The AIPW estimator with $\bLambda$ provided the most reliable estimates of cUHDRS decline in the pre-MCI period (Table~\ref{tab:chp3-hd_app_table}). This estimator indicated a 0.172-point decline in cUHDRS (95\% CI: 0.102–0.242) per year closer to MCI, adjusting for CAG repeat length, education, and sex. Compared to IPW (decline: 0.162, 95\% CI: 0.070–0.254), AIPW with $\bLambda$ achieved improved precision, narrowing the confidence interval by approximately 25\%. The covariate effects were consistent with clinical knowledge: longer CAG repeats and lower education were associated with worse cUHDRS scores. By reliably characterizing how rapidly cUHDRS declines as individuals approach MCI, this improved precision provides a more efficient characterization for trials testing whether interventions can slow functional decline in early disease stages.

In contrast, the MLE performed poorly, potentially mischaracterizing the natural trajectory. Although it estimated a decline of 0.083 points (95\% CI: 0.081–0.085) with the smallest standard errors, the MLE produced counterintuitive covariate effects: greater CAG repeat length and female sex were associated with \emph{higher} cUHDRS scores, contradicting established evidence that longer CAG expansions and female sex are linked to faster decline \citep{schobel2017motor, risby2024sex}. These counterintuitive directions---though not statistically significant---are consistent with the bias patterns observed in our simulations when $f_{X|\bZ}$ is difficult to estimate reliably because $(Y,X,\bZ)$ is rarely observed. Such mischaracterization could lead to flawed assessments of treatment effects in trials using cUHDRS as a primary outcome.

The AIPW estimator using $\bPsi^{\text{eff}}(y,\bz;\btheta)$ also showed instability, estimating a smaller decline of 0.048 points (95\% CI: 0.038–0.058) with counterintuitive sex effects. The counterintuitive effects likely reflect numerical instability in the augmentation term, which involves a ratio of expectations that becomes unstable when the denominator is small.

Our results illustrate the importance of matching estimator choice to censoring rate. Estimators poorly suited to high censoring---whether due to difficulty estimating $f_{X|\bZ}$ (MLE) or numerical instability (AIPW with $\bPsi^{\text{eff}}(y,\bz;\btheta)$)---produced estimates that mischaracterize disease progression, which could lead to incorrect conclusions about treatment efficacy in pre-MCI intervention trials. More broadly, Huntington disease cohorts vary widely in censoring rates depending on follow-up duration and participant characteristics, and applying this framework across studies would improve the reliability and comparability of regression estimates used to contextualize treatment effects.

\begin{table}[h!]
\centering
\resizebox{1\linewidth}{!}{
\begin{tabular}[t]{lccccc}
\toprule
\textbf{Estimator} & \textbf{Intercept} & \textbf{TTMCI (Years)} & \textbf{CAG} & \makecell{\textbf{Education (Years)}} & \textbf{Sex (Female)}  \\
\midrule
\addlinespace
IPW & -1.473 (0.350)$^{**}$ & -0.162 (0.047)$^{**}$ & -0.045 (0.042) & 0.123 (0.051) & -0.086 (0.251) \\
AIPW & -0.998 (0.171)$^{**}$ & -0.048 (0.005)$^{**}$ & -0.022 (0.038) & 0.105 (0.023)$^{**}$ & 0.149 (0.134) \\
AIPW with $\bLambda$ & -1.781 (0.303)$^{**}$ & -0.172 (0.036)$^{**}$ & -0.018 (0.056) & 0.103 (0.051) & -0.050 (0.258) \\
MLE & -1.406 (0.035)$^{**}$ & -0.083 (0.002)$^{**}$ & 0.003 (0.018) & 0.054 (0.024)$^{**}$ & 0.103 (0.090) \\
\bottomrule 
\end{tabular}
}
\caption{\label{tab:chp3-hd_app_table} \textbf{Linear regression parameter estimates (standard errors) of the composite Unified Huntington Disease Rating Scale (cUHDRS), PREDICT-HD ($n=833$).} cUHDRS and Education are centered at their mean; CAP Score is centered at its mean and scaled by a factor of 10; TTMCI refers to Time to MCI, which is calculated as current age minus age at MCI.}
\end{table}

\section{Discussion}
\label{sec:chp3-discussion}

We address a methodological gap in regression with right-censored covariates when censoring depends on the outcome by establishing three consistent estimators---two AIPW estimator variants and the MLE---and deriving their theoretical properties. In our Huntington disease application, the AIPW estimator with $\bLambda$ produced effect estimates aligned with clinical knowledge, outperforming IPW in efficiency and MLE in bias under the high censoring observed in PREDICT-HD.

A key contribution is clarifying how censoring rate should guide estimator choice. Under high censoring, the censoring distribution $f_{C|Y,\bZ}$ can be estimated reliably from the frequently observed triplet $(Y,C,\bZ)$, making weighted-based approaches preferred. In this setting, the AIPW with $\bLambda$ is the to-go estimator: it retains the robustness of the IPW estimator, guarantees efficiency gains over IPW estimator regardless of augmentation term specification, and avoids potentially unstable augmentation terms requiring correct specification of $f_{X|Y,\bZ}$. Under low censoring, $f_{X|\bZ}$ can be estimated reliably from the frequently observed triplet $(Y,X,\bZ)$, making the MLE preferred. Our simulations demonstrated that IPW and AIPW estimators continue to perform well even under low censoring, while MLE performance deteriorates substantially as censoring increases. At moderate censoring rates (around 60\%), all estimators achieved relatively low bias and nominal coverage when correctly specified.

We adopt Weibull AFT models to estimate nuisance parameters and derive the corresponding sandwich variance estimators, though other parametric families or non-parametric approaches may be considered. While IPW has been proposed for outcome-dependent censoring \citep{Lvetal2017, MatsouakaAtem2020}, prior work did not clarify how the sandwich variance estimator must be adjusted to account for variability introduced by estimating weights. We provide the corrected form and demonstrate numerically that the mean estimated standard errors and empirical standard deviations closely agree.

Our simulations confirmed that estimators designed for outcome-independent censoring---including CC, conditional mean imputation, ACC, and MACC---produce biased results under outcome-dependent censoring. Outcome-dependent censoring arises across diverse applications, not just in Huntington disease: policy evaluations modeling unemployment insurance claims as a function of unemployment duration \citep{Lvetal2017} and cardiovascular studies examining cholesterol levels as a function of disease onset \citep{MatsouakaAtem2020},  to name a few. Applying estimators not designed for outcome-dependent censoring in these settings leads to biased effect estimates and potentially incorrect scientific conclusions.

The framework developed here provides practical guidance for researchers facing outcome-dependent censoring: assess the censoring rate, select estimators accordingly, and recognize that the best choice for one study may perform poorly in another with different censoring characteristics. As observational studies increasingly seek to characterize disease trajectories relative to clinical events rather than calendar time, principled approaches to outcome-dependent censoring will be essential for generating reliable disease progression evidence to inform intervention trials.

\section{Software}
\label{sec5}

Software and tutorial are available as an R-package, \textit{\textbf{O}utcome \textbf{D}ependent \textbf{R}ight-\textbf{C}ensored \textbf{C}ovariate (ODRCC)}, is available at \url{https://github.com/jesusepfvazquez/dep-right-censored-covariate}. 

\section{Supplementary Material}
\label{sec6}

Supplementary material is available online at
\url{http://biostatistics.oxfordjournals.org}.

\section*{Acknowledgments}

The authors thank the staff and participants of the PREDICT-HD study for their important contributions. This work was supported by the National Institute of Neurological Disorders and Stroke (R01NS131225; JEV, YM, KM, TPG), Johns Hopkins Provost's Postdoctoral Fellowship Program (JEV), and the National Institute of Environmental Health Sciences (T32ES007018; JEV). {\it Conflict of Interest}: None declared.

\bibliographystyle{biorefs}
\bibliography{biblio}

@string{JRSSC = "Journal of the Royal Statistical Society C"}

@preamble{ "\newcommand{\TPG}{{Garcia, T.P.}}" }

@preamble{ " \newcommand{\noop}[1]{} " }

@article{lee2024robust,
author = {S. Lee and B. D. Richardson and Y. Ma and Karen S. Marder and T. P. Garcia},
title = {SPARCC: Semi-Parametric Robust Estimation in a Right-Censored Covariate Model},
journal = {Journal of the American Statistical Association},
volume = {0},
number = {ja},
pages = {1--22},
year = {2025},
publisher = {ASA Website},
doi = {10.1080/01621459.2025.2562645},
}

@article{AshnerGarcia2023,
  title={Understanding the Implications of a Complete Case Analysis for Regression Models with a Right-Censored Covariate},
  author={Ashner, M. C. and Garcia, T. P.},
  journal={The American Statistician},
  year={2024}, 
  volume = "78", 
  number = "3", 
  pages = "335--344"
}

@article{Atemetal2017,
author={Atem, F.D. and Qian, J. and Maye, J. E. and Johnson, K. A. and Betensky, R. A.},
title={Linear Regression with a Randomly Censored Covariate: {A}pplication to an {A}lzheimer's Study},
journal=JRSSC,
year={2017},
volume={66},
number={2},
pages={313-328}
}

@article{Atemetal2019,
author = {Atem, F. D. and Matsouaka,  R. A and Zimmern,  V. E.},
title = {Cox regression model with randomly censored covariates},
journal = {Biometrical Journal},
year = {2019},
volume = {61},
number = {4},
pages={1020-1032}
}

@article{Bartlettetal2014,
author = {Bartlett, J.W. and Carpenter, J.R. and Tilling, K. and Vansteelandt, S.},
title = {Improving upon the efficiency of complete case analysis when covariates are {MNAR}},
journal = {Biostatistics},
year = {2014},
volume = {15},
number = {4},
pages = {719-730}
}

@article{Bernhardtetal2014,
author =  {Bernhardt, P. W. and Wang, H. J. and Zhang, D.},
title = {Flexible modeling of survival data with covariates subject to detection limits via multiple imputation},
journal = {Computational Statistics and Data Analysis},
year={2014},
volume = {69},
pages={81-91}
}

@article{Bernhardtetal2015,
author = {Bernhardt, P. W. and Wang, H. J. and Zhang, D.},
title = {Statistical Methods for Generalized Linear Model with Covariates Subject to Detection Limits},
year = {2015},
volume = {7},
journal = {Statistics in Biosciences},
pages = {68-79}
}

@article{Lvetal2017,
author = {Lv, X. and Zhang, G. and Li, Q. and Li, R.},
title = {Maximum weighted likelihood for discrete choice models with a dependently censored covariate},
year = {2017},
journal = {Journal of the Korean Statistical Society},
volume = {46},
number = {1},
pages = {15--27}
}

@article{lotspeich2024making,
  title={Making Sense of Censored Covariates: Statistical Methods for Studies of {H}untington's Disease},
  author={Lotspeich, S. C. and Ashner, M. C. and Vazquez, J. E. and Richardson, B. D. and Grosser, K. F. and Bodek, B. E. and Garcia, T. P.},
  journal={Annual Review of Statistics and Its Application},
  volume={11},
  year={2024},
  publisher={Annual Reviews}
}

@article{liu2023using,
  title={Using the {W}eibull accelerated failure time regression model to predict time to health events},
  author={Liu, E. and Liu, R. Y. and Lim, K.},
  journal={Applied Sciences},
  volume={13},
  number={24},
  pages={13041},
  year={2023},
  publisher={MDPI}
}

@article{MatsouakaAtem2020,
author = {Matsouaka, R. A. and Atem, F. D.},
title = {Regression with a right-censored predictor, using inverse probability weighting methods}, 
journal = {Statistics in Medicine}, 
year = {2020},
volume = {39},
number = {27},
pages = {4001--4015}
}

@article{newey1994large,
  title={Large sample estimation and hypothesis testing},
  author={Newey, W. K. and McFadden, D.},
  journal={Handbook of Econometrics},
  volume={4},
  pages={2111--2245},
  year={1994},
  publisher={Elsevier}
}

@article{Qianetal2018,
author = {Qian, J. and 
Chiou, S.H. and 
Maye, J.E. and 
Atem, F. and Johnson, K.A. and Betensky, R.A.},
title = {Threshold regression to accommodate a censored covariate},
journal = {Biometrics},
volume = {74},
number = {4},
pages = {1261-1270},
year = {2018}
}

@article{rodrigues2019huntington,
  title={Huntington’s disease clinical trials corner: January 2019},
  author={Rodrigues, F. B. and Quinn, L. and Wild, E. J.},
  journal={Journal of Huntington's disease},
  volume={8},
  number={1},
  pages={115--125},
  year={2019},
  publisher={SAGE Publications Sage UK: London, England}
}

@article{risby2024sex,
  title={Sex differences in {H}untington's disease from a neuroinflammation perspective},
  author={Risby-Jones, G. and Lee, J. D. and Woodruff, T. M and Fung, J. N.},
  journal={Frontiers in Neurology},
  volume={15},
  pages={1384480},
  year={2024},
  publisher={Frontiers Media SA}
}

@article{rotnitzky1997analysis,
  title={Analysis of semi-parametric regression models with non-ignorable non-response},
  author={Rotnitzky, A. and Robins, J.},
  journal={Statistics in Medicine},
  volume={16},
  number={1},
  pages={81--102},
  year={1997},
  publisher={Wiley Online Library}
}

@article{schobel2017motor,
  title={Motor, cognitive, and functional declines contribute to a single progressive factor in early {HD}},
  author={Schobel, S. A. and Palermo, G. and Auinger, P. and Long, J. D. and Ma, S. and Khwaja, O. S. and Trundell, D. and Cudkowicz, M. and Hersch, S. and Sampaio, C. and others},
  journal={Neurology},
  volume={89},
  number={24},
  pages={2495--2502},
  year={2017},
  publisher={Lippincott Williams \& Wilkins Hagerstown, MD}
}

@article{stern2002cognitive,
  title={What is cognitive reserve? {T}heory and research application of the reserve concept},
  author={Stern, Y.},
  journal={Journal of the {I}nternational {N}europsychological {S}ociety},
  volume={8},
  number={3},
  pages={448--460},
  year={2002},
  publisher={Cambridge University Press}
}

@article{swindell2009accelerated,
  title={Accelerated failure time models provide a useful statistical framework for aging research},
  author={Swindell, W. R.},
  journal={Experimental {G}erontology},
  volume={44},
  number={3},
  pages={190--200},
  year={2009},
  publisher={Elsevier}
}

@Book{Tsiatis2006,
 author = "Tsiatis, A.A.",
 title = "Semiparametric Theory and Missing Data",
 publisher = "Springer",
 year = "2006",
 address = "New York"
}

@article{vazquez2024establishing,
  title={Establishing the Parallels and Differences Between Right-Censored and Missing Covariates},
  author={Vazquez, J. E. and Ashner, M. C. and Ma, Y. and Marder, K. and Garcia, T. P.},
  journal={arXiv preprint arXiv:2409.04684},
  year={2024}
}

@article{williams2009emotional,
  title={The emotional experiences of family carers in Huntington disease},
  author={Williams, J. K. and Skirton, H. and Paulsen, J. S. and Tripp-Reimer, T. and Jarmon, L. and McGonigal K., Meghan and Birrer, E. and Hennig, B. L. and Honeyford, J.},
  journal={Journal of {A}dvanced {N}ursing},
  volume={65},
  number={4},
  pages={789--798},
  year={2009},
  publisher={Wiley Online Library}
}

\section{Supplementary Material}
\label{sec6}

\end{document}


\title{Supplementary Material to \textit{Robust Estimation under Outcome Dependent Right Censoring in Huntington Disease: Estimators for Low and High Censoring Rates}}
\author{JESUS E. VAZQUEZ$^{\ast,1}$, YANYUAN MA$^{2}$, KAREN MARDER$^{3}$, TANYA P. GARCIA$^{4}$\\[3pt]
\textit{\begin{tabular}{@{}c@{}}
$^{1}$Department of Biostatistics, Johns Hopkins University, Baltimore, MD 21231\\
$^{2}$Department of Statistics, Penn State University, State College, PA 16802\\
$^{3}$Department of Neurology, Columbia University Medical Center, New York, NY 10032\\
$^{4}$Department of Biostatistics, University of North Carolina at Chapel Hill, Chapel Hill, NC 27516
\end{tabular}}
\\[6pt]
}

\markboth%
{J. E. Vazquez and others}
{Supplementary Material}

\maketitle

\section{Introduction}

This supplementary material provides detailed proofs of Propositions 1–2 and Theorems 1–2 presented in the main text, for which we have organized it follows:
\begin{itemize}
    \item \textbf{Proposition 1:} We show that the complete case (CC), conditional mean imputation (CMI), augmented complete case (ACC), and modified augmented complete case (MACC) estimators are not consistent under outcome-dependent censoring.
    \item \textbf{Proposition 2:} Consistency of the inverse probability weighted (IPW) estimator.
    \item \textbf{Theorem 1:} Consistency and asymptotic normality of the augmented inverse probability weighted (AIPW) estimator.
    \item \textbf{Theorem 2:} Consistency and asymptotic normality of the maximum likelihood estimator (MLE).
\end{itemize}

Additionally, we provide the analysis that estimates the $f_{C|Y,\bZ}$ distribution as noted in Section 5. Throughout, subscripts are used to distinguish among conditional densities and expectations. For example, $f_{Y|X,\bZ}(y,x,\bz;\btheta)$ denotes the conditional density of $Y$ given $(X,\bZ)$, evaluated at $(Y=y, X=x, Z=z)$, and $E_{Y|X,Z}(\cdot)$ denotes the corresponding conditional expectation. For any vector or matrix $\bM$, we define $\bM^{\otimes 2} \equiv \bM \bM\trans$.

\section{Proof of Proposition 1}
\label{sec:supp-prop1-proof}

To establish that an estimator $\bPhi_{\text{est}}(\bO_i; \btheta)$ is not consistent, it suffices to show that the expectation of its estimating function does not equal zero when evaluated at the true parameter value $\btheta_0$, as this indicates that the estimating equation is biased \citep{Tsiatis2006}---i.e., 
$E_{Y,W,\Delta,\bZ}\{\bPhi_{\text{est}}(\bO_i; \btheta_0)\}\ne\bzero$. We show that the expectation of the estimating equations for the CC, conditional mean imputation, ACC, and MACC estimators are nonzero under outcome-dependent censoring, establishing that these estimators are not consistent in this setting.

\begin{enumerate}

\item \textbf{CC estimator} 

The complete case estimator is of the form,
\bse
\sumi \bPhi_{\rm CC}(\bO_i; \btheta) = \sumi \delta_i \bS_{\btheta}^F (y_i,w_i,\bz_i) = \bzero.
\ese
To prove that the estimator is not consistent, we will show that the expectation of the estimating equation is not equal to zero. More specifically,
\bse
E_{Y,W,\Delta,\bZ} \left\{ \bPhi_{\rm CC}(\bO; \btheta) \right\} &\equiv & E_{Y,W,\Delta,\bZ} \left\{ \Delta \bS_{\btheta}^F (Y,W,\bZ) \right\}  \\
&=&  \int \delta \bS_{\btheta}^F (y,w,\bz) f_{Y,W,\Delta, \bZ}(y,w,\delta,\bz) dy dw d\delta dz \\
&=&  \int I(x \leq c) \bS_{\btheta}^F (y,x,\bz) f_{Y,X,C,\bZ}(y,x,c,\bz) dy dx dc dz \\
&=&  \int I(x \leq c) \bS_{\btheta}^F (y,x,\bz) f_{C|Y,X,\bZ}(c,y,x,\bz) f_{Y,X,\bZ}(y,x,\bz) dy dx dc dz \\
&=&  \int \pi_{Y,X,\bZ}(y,x,\bz) \bS_{\btheta}^F (y,x,\bz) f_{Y|X,\bZ}(y,x,\bz) f_{X,\bZ}(x,\bz) dy dx dz \\
&=&  E_{Y,X,\bZ} \left\{ \pi_{Y,X,\bZ}(Y,X,\bZ) \bS_{\btheta}^F (Y,X,\bZ) \right\}
\ese
We are left with $E_{Y|X,\bZ} \left\{ \pi_{Y,X,\bZ}(Y,x,\bz) \bS_{\btheta}^F (Y,x,\bz) \right\} $, which is not guaranteed to be equal to zero as previously noted by \cite{Lvetal2017}. Thus the CC estimator is not consistent. 

\item \textbf{Conditional mean imputation estimator}

\noindent \textbf{Case 1: $E_{X|Y,\bZ,\Delta=0}(X)$}

The conditional mean imputation estimator is similar to the CC estimator but replaces right-censored values of  $X$  with their imputed counterparts. The form of the conditional mean imputation estimator is given by
\bse
\sumi \bPhi_{\rm CMI}(\bO_i; \btheta) = \sumi \delta_i \bS_{\btheta}^F (y_i,w_i,\bz_i) + (1 - \delta_i) \bS_{\btheta}^F \{y,E_{X|Y,\bZ,\Delta=0}(X),\bz\} = \bzero.
\ese
The first part is the CC estimator, and the second component is the score equation that replaces the value of $X$ with  $E_{X|Y,\bZ,\Delta=0}(X)$.  This conditional mean is equal to
\bse
 E_{X|Y, \bZ, \Delta = 0}(X)  &=& \int x f_{X|Y, \bZ, \Delta = 0} (x,y,\bz, \delta) dx \\
 &=& \frac{\int_{c}^{\infty} x  f_{X \mid Y, \bZ} (x, y, \bz) dx}{\int_c^\infty  f_{X \mid Y, \bZ} (x, y, \bz) dx } \\
&=& \frac{\int_{c}^{\infty} x  f_{X \mid Y, \bZ} (x, y, \bz) dx}{S_{X|Y,\bZ}(c, y,\bz) } \\
&=& c  + S_{X|Y,\bZ}(c, y,\bz)^{-1} \int_{c<x} S_{X \mid Y,\bZ} (x,y,\bz) dx,
\ese
where $S_{X \mid Y,\bZ} (c,y,\bz) = \int_c^\infty  f_{X \mid Y, \bZ} (x, y, \bz) dx$. The last line follows by using integration by parts, since 
\bse
&& \int_{c}^{\infty} x f_{X \mid Y, \bZ} (x, y,\bz) dx \\
&& = \left\{ \lim_{t \rightarrow \infty}  - t S_{X \mid Y,\bZ} (t,y,\bz) \right\}  - \left\{ \lim_{t \rightarrow c} - t S_{X \mid Y,\bZ} (t,y,\bz) \right\} + \int_{c<x} S_{X \mid Y,\bZ} (x,y,\bz) dx  \\
&& =  c S_{X \mid Y,\bZ} (c,y,\bz)  + \int_{c<x} S_{X \mid Y,\bZ} (x,y,\bz) dx.
\ese
Intuitively, this expectation indicates that when $X$ is right-censored by $C$, the conditional mean of $X$ is equal to $C$ plus a positive term, for which we call the \textit{residual life}. \cite{Bernhardtetal2015}, for the limit of detection problem, proved that the conditional mean imputation estimator is not consistent when $X$ depends on $Y$. Here, we show that this bias persists for the right-censored covariate problem because since $Y$ is a function of $(X,\bZ)$, thus  $X \notindependent Y$. In other words, $X$ is part of the mean function of $Y$. 

For the simple case that $m(X,\bZ;\bbeta)$ is assumed to be linear, the conditional mean imputation estimator can also be expressed as,
\bse
\sumi \bPhi_{\rm CMI}(\bO_i; \btheta) = \sumi \bS_{\btheta}^F\left[y,  \{ \delta w + (1-\delta) E_{X|Y,\bZ,\Delta=0}(X) \}, \bz\right] = \bzero.
\ese
Let $\dot{\theta_k} l(y,w,z; \btheta) = \partial \log f_{Y|X,\bZ} (y,w,\bz;\btheta)/ \partial \theta_k$. To improve readability, let $x^* = E_{X|Y,\bZ,\Delta=0}(X)$. The components of $\bS_{\btheta}^F\left[y,  \{ \delta w + (1-\delta) x^*\}, \bz\right]^T$  are equal to 
\bse
\dot{\theta_0} l[y, \{ \delta w + (1-\delta) x^*\},z; \btheta] &=& \epsilon_* /\sigma^2 \\ 
\quad \dot{\beta_1} l[y, \{ \delta w + (1-\delta) x^*\},z; \btheta] &=& \{ \delta w + (1-\delta) x^*\}\epsilon_*/\sigma^2; \\
\dot{\beta_2} l[y, \{ \delta w + (1-\delta) x^*\},z; \btheta] &=& z\epsilon_*/\sigma^2 \\ 
\quad \dot{\sigma} l[y, \{ \delta w + (1-\delta) x^*\},z; \btheta] &=& -1/\sigma + \epsilon_*^2/\sigma^3,
\ese
where $  \epsilon_* = Y - [\beta_0 + \beta_1 \left\{ \Delta W + (1-\Delta) X^* \right\} + \bbeta_2 \bZ]$. We now show that the conditional mean imputation estimator is not consistent by demonstrating that at least one component of $\bS_{\btheta}^F\left[y,  \{ \delta w + (1-\delta) x^*\}, \bz\right]^T$ is not zero in expectation. It follows that
\bse
&& E_{Y,W,\Delta,\bZ} \left(\dot{\theta_0} l[Y, \{ \Delta W + (1-\Delta) X^*\}, \bZ; \btheta] \right) \\ 
&& = \frac{1}{\sigma^2} E_{Y,W,\Delta,\bZ} \left[ Y - (\beta_0 + \bbeta_2 \bZ)  - \beta_1 \left\{ \Delta W + (1-\Delta) X^* \right\}\right] \\
&& = \frac{1}{\sigma^2} E_{Y,\bZ} \left[ Y - (\beta_0 + \bbeta_2 \bZ)  - \beta_1 E_{W,\Delta | Y, \bZ}\left\{ \Delta W + (1-\Delta) X^* \right\}\right] 
\ese
if $E_{W,\Delta | Y, \bZ}\left\{ \Delta W + (1-\Delta) E_{X|Y, \bZ, \Delta = 0} (X) \right\} = E_{X|Y,\bZ}(X)$, then the above is equivalent to the oracle estimator which leads to an expectation equal to zero. It follows that
\bse
&& E_{W,\Delta | Y, \bZ}\left\{ \Delta W + (1-\Delta) E_{X|Y, \bZ, \Delta = 0} (X)  \right\} \\
&& = \int \left\{ \delta w + (1-\delta) E_{X|Y, \bZ, \Delta = 0} (X) \right\} f_{W,\Delta|Y,\bZ} (w,\delta,y, \bz) dw d\delta \\
&& = \int \left\{ \delta w + (1-\delta) E_{X|Y, \bZ, \Delta = 0} (X) \right\} \\
&& \quad \quad \times \left\{ \int_{x \leq C} f_{X,C|Y,\bZ} (w,c,y, \bz) dc \right\}^{\delta} \left\{ \int_{c<X} f_{X,C|Y,\bZ} (x,w,y, \bz) dx \right\}^{1-\delta} dw d\delta \\
&& = \int I(x \leq c ) x  f_{X,C|Y,\bZ} (x,c,y, \bz) dx dc  + \int I(c < x )  E_{X|Y, \bZ, \Delta = 0} (X)  f_{X,C|Y,\bZ} (x,c,y, \bz) dx dc \\
&& = \int \left\{ \int I(x \leq c ) x  f_{X|Y,\bZ} (x,y, \bz) dx \right\} f_{C|Y,\bZ} (x,y, \bz) dc  \\
&& \quad \quad + \int S_{X \mid Y,\bZ} (c,y,\bz)  E_{X|Y, \bZ, \Delta = 0} (X)  f_{C|Y,\bZ} (c,y, \bz) dc.
\ese
Using integration by parts, it follows that 
\bse
&& \int I(x \leq c ) x  f_{X|Y,\bZ} (x,y, \bz) dx \\
&& = \left\{ \lim_{t \rightarrow c}  - t S_{X \mid Y,\bZ} (t,y,\bz) \right\}  - \left\{ \lim_{t \rightarrow \infty} - t S_{X \mid Y,\bZ} (t,y,\bz) \right\} + \int_{x \leq c} S_{X \mid Y,\bZ} (x,y,\bz) dx  \\
&& =  - c S_{X \mid Y,\bZ} (c,y,\bz)  + \int_{x \leq c} S_{X \mid Y,\bZ} (x,y,\bz) dx.
\ese
Therefore, we can reduce the expression as  
\bse
&& E_{W,\Delta | Y, \bZ}\left\{ \Delta W + (1-\Delta) E_{X|Y, \bZ, \Delta = 0} (X)  \right\} \\
&& = \int \left\{ \int I(x \leq c ) x  f_{X|Y,\bZ} (x,y, \bz) dx \right\} f_{C|Y,\bZ} (x,y, \bz) dc  \\
&& \quad \quad \quad + \int S_{X \mid Y,\bZ} (c,y,\bz)  E_{X|Y, \bZ, \Delta = 0} (X)  f_{C|Y,\bZ} (c,y, \bz) dc \\
&& = \int \biggr[ -c S_{X \mid Y,\bZ} (c,y,\bz)  + \biggr\{ \int_{x \leq c}  S_{X \mid Y,\bZ} (x,y,\bz) dx \biggr\}   \\
&& \quad \quad \quad + S_{X \mid Y,\bZ} (c,y,\bz) \biggr\{ c  + S_{X|Y,\bZ}(c, y,\bz)^{-1} \int_{c<x} S_{X \mid Y,\bZ} (x,y,\bz) dx \biggr\} \biggr] f_{C|Y,\bZ} (c,y, \bz) dc \\
&& = \int_{x\leq c} S_{X \mid Y,\bZ} (x,y,\bz) f_{C|Y,\bZ} (c,y, \bz) dc + \int_{c<x} S_{X \mid Y,\bZ} (x,y,\bz) f_{C|Y,\bZ} (c,y, \bz) dc \\
&& = \int S_{X \mid Y,\bZ} (x,y,\bz) f_{C|Y,\bZ} (c,y, \bz) dc \\
 && = S_{X \mid Y,\bZ} (x,y,\bz).
\ese
Finally, since 
\bse
E_{X|Y,\bZ}(X) &=& \int x f_{X|Y,\bZ} (x,y,\bz) dx \\
&\ne& 1 - \pr(X > x | Y=y, \bZ=\bz) \\
&=& S_{X \mid Y,\bZ} (x,y,\bz) 
\ese
then  $E_{Y,W,\Delta,\bZ} \left(\dot{\theta_0} l[Y, \{ \Delta W + (1-\Delta) X^*\}, \bZ; \btheta] \right)$ is not equal to zero. Therefore, the expectation of the estimating equation is not equal to zero,  implying that the conditional mean imputation estimator is not consistent.

\noindent \textbf{Case 2: $E_{X|\bZ}(X)$}

For the simple case that $m(X,\bZ;\bbeta)$ is assumed to be linear in $X$, the conditional mean imputation estimator can also be expressed as
\bse
\sumi \bPhi_{\rm CMI}(\bO_i; \btheta) = \sumi \bS_{\btheta}^F\left[y,  \{ \delta w + (1-\delta) E_{X|\bZ}(X) \}, \bz\right] = \bzero,
\ese
where instead of $E_{X|Y,\bZ,\Delta=0} (X)$, we use $E_{X|\bZ} (X)$. Here, we show that this form of the conditional mean imputation will lead to bias by showing that at least one of the individual components of the estimating equation is not equal to zero in expectation. If follows,
\bse
&& E_{Y,W,\Delta,\bZ} \left(\dot{\theta_0} l[y, \{ \delta x + (1-\delta) E_{X|\bZ} (X)\},\bz; \btheta] \right) \\ 
&& = \frac{1}{\sigma^2} E_{Y,W,\Delta,\bZ} \left[ Y - (\beta_0 + \bbeta_2 \bZ)  - \beta_1 \left\{ \Delta W + (1-\Delta) E_{X|\bZ} (X) \right\}\right] \\
&& = \frac{1}{\sigma^2} E_{Y,\bZ} \left[ Y - (\beta_0 + \bbeta_2 \bZ)  - \beta_1 E_{W,\Delta | Y, \bZ}\left\{ \Delta W + (1-\Delta) E_{X|\bZ} (X) \right\}\right] 
\ese
if $E_{W,\Delta | Y, \bZ}\left\{ \Delta W + (1-\Delta) E_{X|\bZ} (X) \right\} = E_{X|Y,\bZ}(X)$, then the presentation is equivalent to the oracle estimator which leads to an expectation equal to zero. It follows that
\bse
&& E_{W,\Delta | Y, \bZ}\left\{ \Delta W + (1-\Delta) E_{X|\bZ} (X) \right\} \\
&& = \int \left\{ \delta w + (1-\delta) E_{X|\bZ} (X) \right\} f_{W,\Delta|Y,\bZ} (w,\delta,y, \bz) dw d\delta \\
&& = \int \left\{ \delta w + (1-\delta) E_{X|\bZ} (X) \right\} \\
&& \quad \quad \times \left\{ \int_{x \leq C} f_{X,C|Y,\bZ} (w,c,y, \bz) dc \right\}^{\delta} \left\{ \int_{c<X} f_{X,C|Y,\bZ} (x,w,y, \bz) dx \right\}^{1-\delta} dw d\delta \\
&& = \int I(x \leq c ) x  f_{X,C|Y,\bZ} (x,c,y, \bz) dx dc  + \int I(c < x )  E_{X|\bZ} (X)  f_{X,C|Y,\bZ} (x,c,y, \bz) dx dc \\
&& = \int  \pi_{Y,X,\bZ}(y,x,\bz) \ x f_{X|Y,\bZ} (x,y, \bz) dx  + \int \left\{ 1 -  \pi_{Y,X,\bZ}(y,x,\bz) \right\}  E_{X|\bZ} (X)  f_{X|Y,\bZ} (x,y, \bz) dx \\
&& = E_{X|Y,\bZ} \left\{ \pi_{Y,X,\bZ}(y,X,\bz) X \right\} +  E_{X|Y,\bZ} \left[ \left\{ 1 - \pi_{Y,X,\bZ}(y,X,\bz) \right\} E_{X|\bZ}(X) \right].
\ese
The above equation is equal to $E_{X|Y,\bZ}(X)$ if  $E_{X|Y,\bZ} \left[ \left\{ 1 - \pi_{Y,X,\bZ}(y,X,\bz) \right\} E_{X|\bZ}(X) \right] = E_{X|Y,\bZ} \left[ \left\{ 1 - \pi_{Y,X,\bZ}(y,X,\bz) \right\} X \right]$, however, this is not true. Therefore, this form of the conditional mean imputation is also not consistent.  

\item \textbf{ACC Estimator}

The ACC estimator as described in \cite{vazquez2024establishing} is of the form
\bse
\sumi \bPhi_{\rm ACC}(\bO_i; \btheta) = \sumi \delta_i \bS_{\btheta}^F (y_i,w_i,\bz_i) + \left\{\delta_i - \pi_{Y,\bZ}(y_i,\bz_i) \right\} \bPsi_{\rm ACC}(y_i,\bz_i; \btheta) = \bzero,
\ese
where $\pi_{Y,\bZ}(y,\bz) = \int \pi_{Y,X,\bZ}(y,x,\bz) dx$. To show that the ACC is not consistent it is sufficient to show that the expectation of the estimating equation is not equal to $\bzero$. The augmented component is equal to $\bzero$ since, 
\bse
&&  E_{Y,W,\Delta,\bZ}\left[ \left\{\Delta - \pi_{Y,\bZ}(Y,\bZ) \right\} \bPsi_{\rm ACC}(Y,\bZ; \btheta) \right] \\
&&  E_{Y,\bZ} \left[ \left\{ E_{W,\Delta|Y,\bZ} (\Delta) - \pi_{Y,\bZ}(Y,\bZ) \right\} \bPsi_{\rm ACC}(Y,\bZ; \btheta) \right] \\
&&  E_{Y,\bZ} \left[ \left\{ \pi_{Y,\bZ}(Y,\bZ) - \pi_{Y,\bZ}(Y,\bZ) \right\} \bPsi_{\rm ACC}(Y,\bZ; \btheta) \right]  = \bzero.
\ese
Now it is sufficient to show that the component related to the complete case estimator is zero in expectation to show that the ACC estimator is consistent, however, it was previously shown that $E_{Y,W,\Delta,\bZ} \left\{ \Delta \bS_{\btheta}^F (Y,W,\bZ) \right\} \ne \bzero$. This means that  $E_{Y,W,\Delta,\bZ} \left\{ \bPhi_{\rm ACC}(\bO; \btheta) \right\} \ne \bzero$, thus the ACC is not a consistent estimator. Alternatively, we can consider
\bse
\sumi \bPhi^{\rm alt}_{\rm ACC}(\bO_i; \btheta) = \sumi \delta_i \bS_{\btheta}^F (y_i,w_i,\bz_i) + \left\{\delta_i - \pi_{Y,X,\bZ}(y_i,w_i,\bz_i) \right\} \bPsi_{\rm ACC}(y_i,\bz_i; \btheta) = \bzero,
\ese
where instead of $\pi_{Y,\bZ}(y_i,\bz_i)$ we use $\pi_{Y,X,\bZ}(y_i,w_i,\bz_i)$. For this alternative version of the ACC to be consistent, it must follow that  
\bse
&& E_{Y,W,\Delta,\bZ}\left[ \left\{\Delta - \pi_{Y,X,\bZ}(Y,W,\bZ) \right\} \bPsi_{\rm ACC}(Y,\bZ; \btheta) \right] \\
&& \quad \quad = - E_{Y,W,\Delta,\bZ} \left\{ \Delta \bS_{\btheta}^F (Y,W,\bZ) \right\} \\
&& \quad \quad = - E_{Y,X,\bZ} \left\{ \pi_{Y,X,\bZ}(Y,X,\bZ) \bS_{\btheta}^F (Y,X,\bZ) \right\},
\ese
as this result would guarantee $E_{Y,W,\Delta,\bZ} \left\{ \bPhi_{\rm ACC}(\bO; \btheta) \right\} = \bzero$. The augmented component becomes
\bse
&&  E_{Y,W,\Delta,\bZ}\left[ \left\{\Delta - \pi_{Y,X,\bZ}(Y,W,\bZ) \right\} \bPsi_{\rm ACC}(Y,\bZ; \btheta) \right] \\
&& =  \int \left\{ \delta - \pi_{Y,X,\bZ}(y,w,\bz) \right\} \bPsi_{\rm ACC}(y,\bz; \btheta) f_{Y,W,\Delta,\bZ}(y,w,\delta,\bz) dy dw d\delta d\bz \\
&& =  \int I(x \leq c) \left\{1 - \pi_{Y,X,\bZ}(y,x,\bz) \right\} \bPsi_{\rm ACC}(y,\bz; \btheta) f_{Y,X,C,\bZ}(y,x,c,\bz) dy dx dc d\bz \\
&& \quad +  \int I(x > c) \left\{0 - \pi_{Y,X,\bZ}(y,c,\bz) \right\} \bPsi_{\rm ACC}(y,\bz; \btheta) f_{Y,X,C,\bZ}(y,x,c,\bz) dy dx dc d\bz \\
&& =  \int \pi_{Y,X,\bZ}(y,x,\bz) \left\{1 - \pi_{Y,X,\bZ}(y,x,\bz) \right\} \bPsi_{\rm ACC}(y,\bz; \btheta) f_{Y,X,\bZ}(y,x,\bz) dy dx d\bz \\
&& \quad -  \int I(x > c) f_{X|Y,\bZ} (x,y,\bz) dx \ \pi_{Y,X,\bZ}(y,c,\bz) \bPsi_{\rm ACC}(y,\bz; \btheta) f_{Y,C,\bZ}(y,c,\bz) dy dc d\bz,
\ese
where we can express $S_{X|Y,\bZ}(c,y,\bz) = \int I(x > c) f_{X|Y,\bZ} (x,y,\bz) dx$. This leads to the representation
\bse
&&  E_{Y,W,\Delta,\bZ}\left[ \left\{\Delta - \pi_{Y,X,\bZ}(Y,W,\bZ) \right\} \bPsi_{\rm ACC}(Y,\bZ; \btheta) \right] \\
&& =  \int \pi_{Y,X,\bZ}(y,x,\bz) \left\{1 - \pi_{Y,X,\bZ}(y,x,\bz) \right\} \bPsi_{\rm ACC}(y,\bz; \btheta) f_{Y,X,\bZ}(y,x,\bz) dy dx d\bz \\
&& \quad -  \int S_{X|Y,\bZ}(c,y,\bz) \pi_{Y,X,\bZ}(y,c,\bz) \bPsi_{\rm ACC}(y,\bz; \btheta) f_{Y,C,\bZ}(y,c,\bz) dy dc d\bz \\
&& =  E_{Y,X,\bZ} \left[ \pi_{Y,X,\bZ}(Y,X,\bZ) \left\{1 - \pi_{Y,X,\bZ}(Y,X,\bZ) \right\} \bPsi_{\rm ACC}(Y,\bZ; \btheta) \right] \\
&& \quad -  E_{Y,C,\bZ} \left\{ S_{X|Y,\bZ}(C,Y,\bZ) \pi_{Y,X,\bZ}(Y,C,\bZ) \bPsi_{\rm ACC}(Y,\bZ; \btheta) \right\}
\ese
Since the above representation is not equal to $-E_{Y,X,\bZ} \left\{ \pi_{Y,X,\bZ}(Y,X,\bZ) \bS_{\btheta}^F (Y,X,\bZ) \right\}$, then $E_{Y,W,\Delta,\bZ} \left\{ \bPhi_{\rm ACC}(\bO; \btheta) \right\} \ne \bzero$. Therefore, the ACC is not a consistent estimator.

\item \textbf{MACC estimator} 

Under specification of the probability $\pi_{Y,X,\bZ}(y,w,\bz)$, the MACC estimator takes the form,
\bse
\sumi \bPhi_{\rm MACC}(\bO_i; \btheta) = \sumi \delta_i \bS_{\btheta}^F (y_i,w_i,\bz_i) + \left\{ 1 - \frac{\delta_i}{\pi_{Y,X,\bZ}(y_i,w_i,\bz_i)} \right\} \bPsi_{\rm MACC}(y_i,\bz_i; \btheta) = \bzero.
\ese
The first term corresponds to the CC estimator, while the second represents the augmentation component. Compared with the AIPW estimator, the augmentation remains unchanged because the same probability $\pi_{Y,X,\bZ}(y,w,\bz)$ is used and $\bPsi_{\rm MACC}(y,\bz; \btheta)$ is an arbitrary augmentation component that depends only on $(Y,\bZ)$. Therefore, the augmentation component has mean zero, as shown in the proof of Theorem 1.

To guarantee that the MACC estimator is consistent, we must show that the estimating equation is zero in expectation. It remains to show that the component of the CC estimator is zero in expectation. However, as previously shown, this is not the case. Therefore, the MACC estimator is not consistent. 

\end{enumerate}

\section{Proof of Proposition 2: Consistency of the inverse probability weighting estimator}



The IPW estimator is of the form,
\bse
\sumi \bPhi_{\rm IPW}(\bO_i; \btheta) = \sumi \frac{\delta_i \bS_{\btheta}^F (y_i,w_i,\bz_i)}{\pi_{Y,X,\bZ}(y_i,w_i,\bz_i)} = \bzero.
\ese
To prove that the estimator is consistent, assuming standard regularity conditions, we will show that the expectation of the estimating equation is equal to zero. Following a similar argument as the CC estimator, we have
\bse
E_{Y,W,\Delta,\bZ} \left\{ \bPhi_{\rm IPW}(\bO; \btheta) \right\} &\equiv & E_{Y,W,\Delta,\bZ} \left\{ \Delta \bS_{\btheta}^F (Y,W,\bZ) / \pi_{Y,X,\bZ}(Y,W,\bZ) \right\}  \\
&=&  E_{Y,X,\bZ} \left\{ \pi_{Y,X,\bZ}(Y,X,\bZ) \bS_{\btheta}^F (Y,X,\bZ) / \pi_{Y,X,\bZ}(Y,X,\bZ) \right\} \\
&=&  E_{Y,X,\bZ} \left\{ \bS_{\btheta}^F (Y,X,\bZ) \right\} \\
&=&  E_{X,\bZ} [E_{Y|X,\bZ} \left\{ \bS_{\btheta}^F (Y,X,\bZ) \right\}] \\
&=&  E_{X,\bZ} (\bzero) = \bzero 
\ese
The last line follows since $E_{Y|X,\bZ} \left\{ \bS_{\btheta}^F (Y,x,\bz) \right\}= \bzero$, for which we can show since 
\be
\bzero &=& \frac{\partial}{\partial \btheta^T} \int  f_{Y|X,\bZ}(y,x,\bz; \btheta_0) dy; \ \ \text{since} \ 1=\int f_{Y|X,\bZ}(y,x,\bz; \btheta_0) dy \nonumber\\
&=& \int \biggr\{ \frac{\partial}{\partial \btheta^T}  \log f_{Y|X,\bZ}(y,x,\bz; \btheta_0) \biggr\}  f_{Y|X,\bZ}(y,x,\bz; \btheta_0) dy \nonumber
\\
&=& \int \bS_{\btheta}(y,x,\bz)  f_{Y|X,\bZ}(y,x,\bz; \btheta_0) dy \nonumber
\\
&=& E_{Y|X,\bZ} \left\{ \bS_{\btheta}^F (Y,x,\bz) \right\}. \label{eqn:proof-zero}
\ee
The second equality holds by standard regularity conditions that allow us to interchange the order of the integral and partial derivative. Since the expectation is equal to zero, we have shown that the estimating equation for the IPW estimator is equal to $\bzero$ in expectation, thus consistent. 

\noindent \textbf{Robustness:} Consider the case that $f_{C|Y,\bZ}$ is incorrectly specified as $f_{C|Y,\bZ}^*$. Let  $\pi^*_{Y,X,\bZ}(y,x,\bz) = \int_{x \leq C} f^*_{C|Y,\bZ} (c,y,\bz) dc$. Then,
\bse
E\left\{\frac{\Delta \bS_{\btheta}^F (Y,W,\bZ)}{ \pi^*_{Y,X,\bZ}(Y,W,\bZ)}\right\}
&=& E_{Y,\bZ} \biggr[ E_{X|Y,\bZ} \left\{\frac{ \pi_{Y,X,\bZ}(Y,X,\bZ)}{ \pi^*_{Y,X,\bZ}(Y,X,\bZ)} \times \bS_{\btheta}^F (Y,X,\bZ)\right\} + E_{C|Y,\bZ} ( \0 ) \biggr] \\
&=& E_{Y,X,\bZ} \biggr\{ \frac{ \pi_{Y,X,\bZ}(Y,X,\bZ)}{ \pi^*_{Y,X,\bZ}(Y,X,\bZ)} \times   \bS_{\btheta}^F (Y,X,\bZ) \biggr\} \\
&\ne&  E_{Y,X,\bZ} \left\{ \bS_{\btheta}^F (Y,X,\bZ) \right\} 
\ese
The last line follows since $\pi_{Y,X,\bZ}(Y,X,\bZ)/ \pi^*_{Y,X,\bZ}(Y,X,\bZ) \ne 1$. Since we are unable to obtain $E_{Y,X,\bZ} \left\{ \bS_{\btheta}^F (Y,X,\bZ) \right\}$, then, the argument used previously no longer holds. Therefore, the IPW estimator is not robust to misspecification of $f_{C|Y,\bZ}$.


\section{Proof of Theorem 1: Consistency and asymptotic normality of the AIPW estimator}

The proof for Theorem 1 is  divided into multiple parts. First, we address the estimation of the parameter set $\boldeta$ governing the nuisance distribution $f_{C \mid Y,\bZ}$. Then, we  prove consistency, and asymptotic normality. Finally we conclude by deriving the augmented term that improves efficiency over the IPW estimator. 

To prove consistency and asymptotic normality of the estimators we assume the following regularity. For an estimator $\bPhi_{\rm est}(\bO; \btheta, \boldeta)$, we assume 

\begin{enumerate}[label=(A\arabic*),ref=(A\arabic*)]
    \item \label{reg:compact}
    $\btheta_0\in\bOmega$ and $\boldeta_0\in\bOmega_{\boldeta}$, where both $\bOmega$ and $\bOmega_{\boldeta}$ are compact.
    \item\label{reg:bounded}
    $E[\sup_{\btheta\in\Omega}\|\bPhi_{\rm est}(\bO;  \btheta, \boldeta)\|_2]<\infty$. 
    \item\label{reg:smooth-A} $E \{ \partial \bPhi_{\rm est}(\bO;  \btheta_0, \boldeta_0)/ \partial \btheta^T \}$ has bounded eigenvalues and is invertible.
    \item\label{reg:smooth-A-alpha}
    $E \{ \partial \bPhi_{\rm est}(\bO;  \btheta_0, \boldeta_0)/ \partial \boldeta^T \}$ has bounded eigenvalues for $\btheta \in \bOmega$.
\end{enumerate}

Throughout, we work to prove that $\wh\btheta$ is a consistent estimator of $\btheta$ by showing its estimating equation is unbiased; i.e., $E \{ \bPhi_{\rm est}(\bO;  \btheta_0; \boldeta_0) \} = \0$. By showing that the estimator is unbiased and assuming standard regularity conditions \ref{reg:compact}-\ref{reg:smooth-A-alpha}, the estimator $\wh\btheta$ is consistent and asymptotically normal by following Theorem 2.6 of \cite{newey1994large}.

The first three conditions, \ref{reg:compact}-\ref{reg:smooth-A} are typically assumed in large sample theory to show consistency and uniqueness of the estimating equation \citep{newey1994large}. Finally, condition \ref{reg:smooth-A-alpha} regulates the influence of the nuisance parameters for which is reasonable since we consider parametric accelerated failure time models to estimate the parameters governing the nuisance distributions. These AFT models are parametric, meaning that they have a finite-dimensional nuisance parameter set, and in our implementation and are sufficiently smooth. Additionally, we assume that all probabilities are bounded (not including) 0 and 1, which is an important assumption for the weighted estimators as they are weighted inversely proportional to this probability. 

\subsection{Estimation of nuisance distributions}
\label{sec:chp3-est_nuisance_distributions}  

Estimation of $\btheta$ when using the AIPW estimator requires specifying $f_{C|Y,\bZ}$. Throughout, we consider a parametric model for $f_{C|Y,\bZ}$, where $\boldeta$ is the set of finite-dimensional parameters; we refer to them as nuisance parameters. The log-likelihood for one individual is given by,
\be
&& \log f_{Y,W,\Delta,\bZ}(y_i, w_i,\delta_i, \bz_i) \nonumber \\
&& = \delta_i \log\left\{ \int_{w_i\leq C} f_{Y,C,X,\bZ}(y_i,w_i,c,\bz_i) dc \right\} + (1-\delta_i) \left\{ \int_{w_i\leq X} f_{Y,C,X,\bZ}(y_i,x,w_i,\bz_i) dx \right\} \nonumber  \\
&& = \delta_i \log\left\{ f_{X|Y,\bZ}(w,y_i,\bz_i) \int_{w_i\leq C} f_{C|Y,X,\bZ}(c,y_i,w_i,\bz_i) dc \right\} + \nonumber \\
&& + (1-\delta_i) \log \left\{ \int_{w_i\leq X} f_{X|Y,\bZ}(x,y_i,\bz_i) f_{C|Y,X,\bZ}(w_i,y_i,x,\bz_i)  dx \right\} + \log f_{Y,\bZ}(y_i, \bz_i) \nonumber \\
&& = \delta_i \log\left\{ f_{X|Y,\bZ}(w,y_i,\bz_i) \int_{w_i\leq C} f_{C|Y,\bZ}(c,y_i,\bz_i; \boldeta) dc \right\} + \nonumber \\
&& + (1-\delta_i) \log \left\{ \int_{w_i\leq X} f_{X|Y,\bZ}(x,y_i,\bz_i) f_{C|Y,\bZ}(w_i,y_i,\bz_i; \boldeta)  dx \right\} + \log f_{Y,\bZ}(y_i, \bz_i).  \label{eqn:log-likelihood-xc-yz}
\ee
The last equality follows since $X \independent C | (Y,\bZ)$. Then, for the density $f_{C|Y, \bZ}(c,y,\bz; \boldeta)$, the score equation is equal to
    \bse
     \bS_{\boldeta}(w_i,\delta_i,y_i,\bz_i; \boldeta) &=& \frac{\partial}{\partial \boldeta^T} \log f_{Y,W,\Delta,\bZ}(y_i, w_i,\delta_i, \bz_i) \\
     &=& \frac{\partial}{\partial \boldeta^T} \left\{ \delta_i \log \int_{w\leq C} f_{C|Y,\bZ}( c,y,\bz;\boldeta) dc + (1-\delta_i) \log f_{ C|Y,\bZ}( c,y,\bz;\boldeta) \right\} \\
     &=& \delta_i \frac{\int_{w_i\leq C} \{ \partial \log f_{C|Y,\bZ}(c,y_i,\bz_i; \boldeta) /\partial \boldeta^T  \}  f_{C|Y,\bZ}(c,y_i,\bz_i; \boldeta) dc}{\int_{w_i\leq C} f_{C|Y,\bZ}(c,y_i,\bz_i; \boldeta) dc} \\
    && + (1-\delta_i)  \partial \log f_{C|Y,\bZ}(w_i,y_i,\bz_i; \boldeta) /\partial \boldeta^T. 
    \ese
    Finally, let $\wh\boldeta$ be the solution to the estimating equation 
    \be
    \label{eqn:chp3-eta_est}
     \sumi \bPhi_{\boldeta}(\bO_i; \boldeta) = \sumi \bS_{\boldeta}(w_i,\delta_i,y_i,\bz_i; \boldeta) = \bzero.
    \ee
     The density $f_{C|Y,\bZ}$ follows a Weibull distribution based on the specification of the AFT Weibull regression model and $\log(C)$ follows a Gumbel distribution. Without loss of generality, let $C = \log(C)$ and $W = \log(W)$. Then it follows that
     \bse
        f_{C|Y,\bZ}(c,y,\bz;\boldeta) &=& \frac{1}{\tilde\sigma_\boldeta} \exp\left\{ \frac{c - m_{\tilde{\boldeta}}(y,\bz;\tilde\boldeta)}{\tilde\sigma_\boldeta} \right\} \exp \left[ - \exp\left\{ \frac{c - m_{\tilde{\boldeta}}(y,\bz;\tilde\boldeta)}{\tilde\sigma_\boldeta} \right\} \right] \\
        \log f_{C|Y,\bZ}(c,y,\bz;\boldeta) &=& -\log \  \tilde\sigma_\boldeta +  \frac{c - m_{\tilde{\boldeta}}(y,\bz;\tilde\boldeta)}{\tilde\sigma_\boldeta} - \exp\left\{ \frac{c - m_{\tilde{\boldeta}}(y,\bz;\tilde\boldeta)}{\tilde\sigma_\boldeta} \right\}.  
     \ese

     Let $\tilde e_i = c - m_{\tilde\boldeta} (y_i,\bz_i; \tilde\boldeta)$. Suppose that $m_{\tilde{\boldeta}}(y,\bz;\tilde\boldeta)$ is of the linear form 
     \bse
     m_{\tilde{\boldeta}}(y,\bz;\tilde\boldeta) = \tilde\eta_0 + \tilde \eta_1 y  + \tilde\eta_2z_1 + ... \tilde\eta_{p+1} z_p.
     \ese
     Then each of the individual components of $\partial \log f_{C|Y,\bZ}(w_i,y_i,\bz_i; \boldeta) /\partial \boldeta^T$ is of the form
     \bse
     && \frac{\partial }{\partial \boldeta^T}  \log f_{C|Y,\bZ}(w_i,y_i,\bz_i; \boldeta)  \\
    && =
    \begin{bmatrix}
        \partial \log f_{C|Y,\bZ}(w_i,y_i,\bz_i; \boldeta) /\partial \tilde\sigma_\boldeta \\
        \partial \log f_{C|Y,\bZ}(w_i,y_i,\bz_i; \boldeta) /\partial \tilde\eta_0 \\
        ... \\
        \partial \log f_{C|Y,\bZ}(w_i,y_i,\bz_i; \boldeta) /\partial \tilde\eta_{p+1} \\
    \end{bmatrix}
    =
    \begin{bmatrix}
        -1/\tilde\sigma_{\boldeta} - \tilde e_i / \tilde\sigma_{\boldeta}^2 \times [1 - \exp\{\tilde e_i/\tilde\sigma_{\boldeta} \}] \\
         -1/\tilde\sigma_{\boldeta} \times [1 - \exp\{\tilde e_i/\tilde\sigma_{\boldeta} \}]  \\
        ... \\
        -z_p/\tilde\sigma_{\boldeta} \times [1 - \exp\{\tilde e_i/\tilde\sigma_{\boldeta} \}] 
    \end{bmatrix}.
 \ese

\subsection{Consistency} {\label{thm1:chp3-consistency}}

In this section we now prove that \eqref{eqn:chp3-eta_est} and that the AIPW estimators are consistent. 

\begin{enumerate}

     \item \textbf{Accelerated failure time} 
     
     $\bS_{\boldeta}(w,\delta,y,\bz;\boldeta_0) = \partial f_{W,\Delta|Y,\bZ}(w,\delta,y,\bz; \boldeta_0) / \partial \boldeta$ where 
    \bse
      f_{W,\Delta|Y,\bZ}(w,\delta,y,\bz; \boldeta_0) &=& \delta_i \log\left\{ f_{X|Y,\bZ}(w,y_i,\bz_i) \int_{w_i\leq C} f_{C|Y,\bZ}(c,y_i,\bz_i; \boldeta) dc \right\} +  \\
&& \quad \quad +  (1-\delta_i) \log \left\{ \int_{w_i\leq X} f_{X|Y,\bZ}(x,y_i,\bz_i) f_{C|Y,\bZ}(w_i,y_i,\bz_i; \boldeta)  dx \right\}.
    \ese
    Since $1=\int f_{W,\Delta|Y,\bZ}(w,\delta,y,\bz; \boldeta_0) dwd\delta$, then it follows that 
    \be
    \bzero &=& \frac{\partial}{\partial \boldeta^T} \int  f_{W,\Delta|Y,\bZ}(w,\delta,y,\bz; \boldeta_0) dc \nonumber\\
    &=& \int \biggr\{ \frac{\partial}{\partial \boldeta^T}  \log f_{W,\Delta|Y,\bZ}(w,\delta,y,\bz; \boldeta_0) \biggr\} f_{W,\Delta|Y,\bZ}(w,\delta,y,\bz; \boldeta_0) dw d\delta \nonumber
    \\
    &=& \int \bS_{\boldeta}(w,\delta,y,\bz;\boldeta_0) f_{W,\Delta|Y,\bZ}(w,\delta,y,\bz; \boldeta_0) dw d\delta \nonumber
    \\
    &=& E_{W,\Delta|Y,\bZ}\{\bS_{\boldeta}(W,\Delta,Y,\bZ;\boldeta_0)\}. \label{eqn:proof-zero}
    \ee
The second equality holds by standard regularity conditions that allow us to interchange the order of the integral and partial derivative. It therefore follows that 
\bse
E_{W,\Delta,Y,\bZ}\{\bS_{\boldeta}(W,\Delta,Y,\bZ;\boldeta_0)\} &=& E_{Y,\bZ} [ E_{W,\Delta|Y,\bZ} \{\bS_{\boldeta}(W,\Delta,Y,\bZ;\boldeta_0)\} ] = \bzero.
\ese

\item \textbf{AIPW estimator}

The AIPW estimator solves the system
\bse
\sumi \bPhi_{\rm AIPW}(\bO_i; \btheta) = \sumi \frac{\delta_i \bS_{\btheta}^F (y_i,w_i,\bz_i)}{\pi_{Y,X,\bZ}(y_i,w_i,\bz_i)} + \left\{ 1 - \frac{\delta_i}{\pi_{Y,X,\bZ}(y_i,w_i,\bz_i)} \right\} \bPsi_{\rm AIPW}(y_i,\bz_i; \btheta) = \bzero.
\ese
To prove that the estimator is consistent, we will show that the expectation of the estimating equation is equal to zero. We have already shown that the IPW estimator is consistent. Now, it remains to show that the augmentation term has mean zero. It follows that
\bse
&& E_{Y,W,\Delta,\bZ} \left[\left\{ 1 - \frac{\Delta}{\pi_{Y,X,\bZ}(Y,W,\bZ)} \right\} \bPsi_{\rm AIPW}(Y,\bZ; \btheta) \right] \\
&& = E_{Y,\bZ} \left(\left[ 1 - E_{W, \Delta|Y,\bZ}\left\{\frac{\Delta}{\pi_{Y,X,\bZ}(Y,W,\bZ)} \right\} \right] \bPsi_{\rm AIPW}(Y,\bZ; \btheta) \right)   \\
&& = E_{Y,\bZ} \left\{ \left( 1 - 1 \right) \bPsi_{\rm AIPW}(Y,\bZ; \btheta) \right\} \\
&& = E_{Y,\bZ} \left\{ 0 \times \bPsi_{\rm AIPW}(Y,\bZ; \btheta) \right\} = \bzero.
\ese
The second to last line follows since
\bse
&& E_{W, \Delta|Y,\bZ}\left\{\frac{\Delta}{\pi_{Y,X,\bZ}(y,W,\bz)} \right\} \\
&& = \int \frac{\delta}{\pi_{Y,X,\bZ}(y,w,\bz)} f_{W,\Delta|Y,\bZ} (w,\delta, y,\bz) dw d\delta \\
&& = \int \frac{\delta}{\pi_{Y,X,\bZ}(y,w,\bz)} \left\{ \int_{x \leq w }  f_{X,C|Y,\bZ} (w,c, y,\bz) dc \right\}^{\delta} \left\{ \int_{c \leq w }  f_{X,C|Y,\bZ} (x,w, y,\bz) dc  \right\}^{1-\delta} dw d\delta \\
&& = \int \frac{I(x \leq c)}{\pi_{Y,X,\bZ}(y,x,\bz)} f_{C|Y,X,\bZ} (c,y,x,\bz) f_{X|Y,\bZ} (x,y,\bz) dx dc \delta \\
&& \stackrel{C \independent Y | (X,\bZ)}{=} \int \frac{I(x \leq c)}{\pi_{Y,X,\bZ}(y,x,\bz)} f_{C|Y,\bZ} (c,y,\bz) dc f_{X|Y,\bZ} (x,y,\bz) dx \\
&& = \int \frac{\pi_{Y,X,\bZ}(y,x,\bz)}{\pi_{Y,X,\bZ}(y,x,\bz)} f_{X|Y,\bZ} (x,y,\bz) dx \\
&& = \int f_{X|Y,\bZ} (x,y,\bz) dx =  1.
\ese
Since the expectation is equal to zero, we have shown that the augmentation term is zero in expectation, thus the AIPW estimator is consistent. 

\textbf{Robustness:} To show that the AIPW estimator is only singly robust, it suffices to show that the estimating equation is unbiased only when the probabilities are correctly specified. We have already proved that the AIPW estimator is consistent when the probabilities are correctly specified. Now, consider the case that $\pi_{Y,X,\bZ}(y,w,\bz)$ is incorrectly specified as $\pi^*_{Y,X,\bZ}(y,w,\bz)$ and $\bPsi_{\rm AIPW} (y,\bz;\btheta) = \bPsi_{\rm AIPW}^{\rm eff} (y,\bz;\btheta)$, where $\bPsi_{\rm AIPW}^{\rm eff} (y,\bz;\btheta) = \frac{ E_{X | Y,\bZ} \left[ \{1-1/\pi_{Y,X,\bZ}(y,X,\bz)\} \bS_{\btheta}^F (y,X,\bz)  \right]}{E_{X | Y,\bZ}  \{1-1/\pi_{Y,X,\bZ}(y,X,\bz)\}}$. We later show that this choice guarantees higher efficiency over the IPW estimator. Then, it follows
\bse
&& E_{Y,W,\Delta,\bZ}[\{1 - \Delta/\pi_{Y,X,\bZ}^*(Y,W,\bZ) \}\bPsi_{\rm AIPW}(Y,\bZ;\btheta)] \\
&& = E_{Y,X,\bZ} \left( \{1 - \pi_{Y,X,\bZ}(Y,X,\bZ)/\pi_{Y,X,\bZ}^*(Y,X,\bZ) \}\bPsi_{\rm AIPW}(Y,\bZ;\btheta)] \right) \\
&& = E_{Y,X,\bZ} \biggr( \left\{ 1 - \frac{\pi_{Y,X,\bZ}(Y,X,\bZ)}{\pi_{Y,X,\bZ}^*(Y,X,\bZ)} \right\} \frac{ E_{X | Y,\bZ} \left[ \{1-1/\pi^*_{Y,X,\bZ}(y,X,\bz)\} \bS_{\btheta}^F (y,X,\bz)  \right]}{E_{X | Y,\bZ}  \{1-1/\pi^*_{Y,X,\bZ}(y,X,\bz)\}} \biggr)\\
&& \ne \bzero,
\ese 
where the last line follows since $\pi_{Y,X,\bZ}(y,x,\bz) \ne \pi_{Y,X,\bZ}^*(y,x,\bz)$. This result shows that the AIPW estimator is consistent only when $\pi_{Y,X,\bZ}(y,x,\bz)$ is correctly specified. Therefore, the AIPW estimator is not robust to the misspecification of $f_{C|Y,\bZ}$.

\end{enumerate}

\subsection{Asymptotic normality} 
\label{thm1:asymptotic-normality}

%
We begin by establishing the asymptotic normality of the consistent estimators using the general theory of $m$-estimation \citep{Tsiatis2006}. By applying a similar argument to that of \cite{vazquez2024establishing} it follows that the distribution of the AIPW estimator, $\wh{\btheta}_{{\rm AIPW, AFT}}$, is
\bse
n^{1/2}(\wh{\btheta}_{{\rm AIPW, AFT}}-\btheta_0) \rightarrow_d \Normal(\bzero, \bA_{\rm AIPW,AFT}^{-1}\bB_{\rm AIPW,AFT}\bA_{\rm AIPW,AFT}^{-T}), 
\ese
for which the components are 
\bse
\bA_{\rm AIPW, AFT} &=& E\left\{ \  \partial  \bPhi_{\rm AIPW}(\bO; \btheta_0, \boldeta_0) / \partial \btheta^T \right\} \\
\bB_{\rm AIPW, AFT} &=&  E\{(\bPhi_{\rm AIPW}(\bO; \btheta_0,\boldeta_0) \\
&& - E\{\partial\bPhi_{\rm AIPW}(\bO;\btheta_0,\boldeta)/\partial\boldeta^T |_{\boldeta = \boldeta_0} \} [E\{\partial\bPhi_{\rm AFT}(\bO;\boldeta)/\partial\boldeta^T |_{\boldeta = \boldeta_0}\}]^{-1} \bPhi_{\rm AFT}(\bO;\boldeta_0) )^{\otimes2}\}.
\ese
Note that $\bA_{\rm AIPW, AFT}$ can be reduced to
\bse
\bA_{\rm AIPW, AFT} &=& E\biggr\{ \frac{\partial}{\partial \btheta^T}    \bPhi_{\rm AIPW}(\bO; \btheta_0, \boldeta_0) \biggr\} \\
&=& E \biggr[   \frac{\partial}{\partial \btheta^T}  \frac{\Delta \bS^F_\btheta(Y,W,\bZ;\btheta_0)}{\pi_{Y,X,\bZ}(Y,W, \bZ; \boldeta_0)} +  \biggr\{1- \frac{\Delta}{\pi_{Y,X,\bZ}(Y,W, \bZ; \boldeta_0)} \biggr\}  \biggr\{ \frac{\partial}{\partial \btheta^T} \bPsi_{\rm AIPW}(Y, \bZ;\btheta_0) \biggr\} \biggr]\\
&=& E \biggr\{   \frac{\partial}{\partial \btheta^T}  \frac{\Delta \bS^F_\btheta(Y,W,\bZ;\btheta_0)}{\pi_{Y,X,\bZ}(Y,W, \bZ; \boldeta_0)} \biggr\}. 
\ese 
It follows that the influence function can be characterized by 
\bse
\bUpsilon_{ \rm AIPW; AFT} (\bO_i) &=& \bA_{\rm AIPW, AFT}^{-1} \{ (\bPhi_{\rm AIPW}(\bO_i; \btheta_0,\boldeta_0) \\
&& \quad \quad - E\{\partial\bPhi_{\rm AIPW}(\bO;\btheta_0,\boldeta)/\partial\boldeta^T |_{\boldeta = \boldeta_0} \}[E\{\partial\bPhi_{\rm AFT}(\bO;\boldeta)/\partial\boldeta^T |_{\boldeta = \boldeta_0}\}]^{-1}\bPhi_{\rm AFT}(\bO_i;\boldeta_0) )\}.
\ese
The term $\bPhi_{\rm AFT}(\bO;\boldeta_0)$ represents the estimating equation \eqref{eqn:chp3-eta_est} corresponding to the nuisance parameters $\boldeta$. The  estimate of the sandwich estimator for the asymptotic variance of $\wh\btheta_{\rm AIPW, AFT}$, under the presence of $\wh\boldeta$, is $\var(\wh\btheta_{\rm AIPW, AFT}) \equiv \var\{ \bUpsilon_{ \rm AIPW, AFT} (\bO) \} = \wh\bA_{\rm AIPW, AFT}^{-1} \wh\bB_{\rm AIPW, AFT} \wh\bA_{\rm AIPW, AFT}^{-T}$, where $\wh\bA_{\rm AIPW, AFT}$ and $\wh\bB_{\rm AIPW, AFT}$ are $\bA_{\rm AIPW, AFT}$ and $\bB_{\rm AIPW, AFT}$ evaluated at $(\wh\btheta_{\rm AIPW, AFT}, \wh\boldeta)$, respectively. 

When \emph{the only unknown parameter set is $\btheta$,} and no estimation of the nuisance parameters is needed, then the form of the asymptotic distribution of $\wh\btheta_{\rm AIPW}$ is simplified. It follows that
\bse
n^{1/2}(\wh\btheta_\text{AIPW} - \btheta_0) \rightarrow_d \Normal (\bzero, \bA_{\rm AIPW}^{-1} \bB_{\rm AIPW} \bA_{\rm AIPW}^{-T} ),
\ese
where $\bA_{\rm AIPW} = E\biggr\{ \frac{\partial}{\partial \btheta^T} \bPhi_{\rm AIPW}(\bO;\btheta_0)\biggr\} $ and $\bB_{\rm AIPW}= E\biggr\{ \bPhi_{\rm AIPW}(\bO;\btheta_0) ^{\otimes 2} \biggr\}$. The corresponding influence function of $\wh\btheta$ is $\bUpsilon_{\rm AIPW}(\bO_i) = - \bA_{\rm AIPW}^{-1} \bPhi_{\rm AIPW}(\bO;\btheta_0)$, which is estimated by substituting $\wh\btheta$ for $\btheta_0$. The sandwich estimator for the asymptotic variance of $\wh\btheta$ is  $\var(\wh\btheta_{\rm AIPW}) = \wh\bA_{\rm AIPW}\wh\bB_{\rm AIPW} \wh\bA_{\rm AIPW}^{-T}$, where $\wh\bA_{\rm AIPW}$ and $\wh\bB_{\rm AIPW}$ are $\bA_{\rm AIPW}$ and $\bB_{\rm AIPW}$ evaluated at $\wh\btheta$, respectively.


When the nuisance parameters indexing the nuisance distribution $f_{C|Y,\bZ}$ are assumed known, the corresponding influence function is 
\bse
\bUpsilon_\text{AIPW}(\bO_i) = - \bA_\text{AIPW}^{-1} \bPhi_\text{AIPW}(\bO;\balpha_0).
\ese
The components of the asymptotic variance are $\bA_{\rm AIPW} = E\{\partial \bPhi_{\rm AIPW}(\bO; \btheta, \boldeta_0)/\partial\btheta^{\trans} |_{\btheta = \btheta_0}\}$ and $\bB_{\rm AIPW} = E\{\bPhi_{\rm AIPW}(\bO;\btheta_0,\boldeta_0)^{\otimes2}\}$. Note that the the influence function of the IPW estimator, as well as it's asymptotic normality and robust sandwich estimator form, can be obtained using a similar argument by replacing $\bPhi_{\rm AIPW}(\bO; \btheta, \boldeta_0)$ with $\bPhi_{\rm IPW}(\bO; \btheta, \boldeta_0)$.


\subsection{Efficiency}
\label{sec:chp3-aipw-efficiency}


Let $\bUpsilon_{\rm IPW}$ and $\bUpsilon_{\rm AUG}$ be the influence functions of the IPW and AIPW estimators, respectively. Then, the AIPW estimator is guaranteed to be more efficient than the IPW estimator under the efficiency condition: 
\be
\label{eqn:chp3-efficiency-condition}
\0 &=& {\rm cov} (\bUpsilon_{\rm AIPW} - \bUpsilon_{\rm IPW}, \bUpsilon_{\rm AIPW}) = \var(\bUpsilon_{\rm AIPW}) - {\rm cov} (\bUpsilon_{\rm AIPW}, \bUpsilon_{\rm IPW}).
\ee
Following the same steps to the AIPW estimator developed in \cite{vazquez2024establishing}, but now with the probability $\pi_{Y,X,\bZ}(y,w,\bz)$ as opposed to $\pi_{X,\bZ}(w,\bz)$, we now show how to apply the efficiency condition \eqref{eqn:chp3-efficiency-condition} to construct the efficient estimating function $\bPsi^{\rm eff}_{\rm AIPW}(y, \bz; \btheta)$. 

For notational simplicity, let $\bPhi_{\rm IPW}(\bO; \btheta)$ and $\bPhi_{\rm AIPW}(\bO; \btheta)$ denote the estimating equations for the IPW and AIPW estimators, respectively. It follows that
\bse
\bzero &=&  \cov(\bUpsilon_{\rm AIPW} - \bUpsilon_{\rm IPW}, \bUpsilon_{\rm AIPW}) \\
&=& \cov(\bUpsilon_{\rm AIPW}) - \cov(\bUpsilon_{\rm IPW}, \bUpsilon_{\rm AIPW}) \\
&=& \bA_{\rm AIPW} E_{Y,W,\Delta,\bZ}\{ \bPhi_{\rm AIPW}(\bO; \btheta) ^{\otimes 2}\} \bA_{\rm AIPW}^T - \bA_{\rm IPW} E_{Y,W,\Delta,\bZ}\{ \bPhi_{\rm IPW}(\bO; \btheta)\bPhi_{\rm AIPW}(\bO; \btheta)^T \} \bA_{\rm AIPW}^T \\
&=& E_{Y,W,\Delta,\bZ}\{ \bPhi_{\rm AIPW}(\bO; \btheta) ^{\otimes 2}\} - E_{Y,W,\Delta,\bZ}\{ \bPhi_{\rm IPW}(\bO; \btheta)\bPhi_{\rm AIPW}(\bO; \btheta)^T \},
\ese
where, $\bA_{\rm AIPW} = \bA_{\rm IPW}$. Then it follows that 
\bse
&& E_{Y,W,\Delta,\bZ}\{ \bPhi_{\rm AIPW}(\bO; \btheta) ^{\otimes 2}\} \\
&& = E_{Y,W,\Delta,\bZ}\{ \bPhi_{\rm IPW}(\bO; \btheta) ^{\otimes 2}\} + E_{Y,W,\Delta,\bZ} [ \bPhi_{\rm IPW}(\bO; \btheta) \{1-\Delta/\pi_{Y,X,\bZ}(Y,W,\bZ)\}\bPsi_{\rm AIPW}(Y,\bZ; \btheta)^T ] \\
&& \quad  + E_{Y,W,\Delta,\bZ}[ \{1-\Delta/\pi_{Y,X,\bZ}(Y,W,\bZ)\} \bPsi_{\rm AIPW}(Y,\bZ; \btheta) \bPhi_{\rm IPW}(\bO; \btheta)^T ] \\
&& \quad + E_{Y,W,\Delta,\bZ} [ \{1-\Delta/\pi_{Y,X,\bZ}(Y,W,\bZ)\}^2 \bPsi_{\rm AIPW}(Y,\bZ; \btheta)^{\otimes 2}]  
\ese
The second part of the equation is equal to
\bse
&& E_{Y,W,\Delta,\bZ}\{ \bPhi_{\rm IPW}(\bO; \btheta)\bPhi_{\rm AIPW}(\bO; \btheta)^T \} \\
&& = E_{Y,W,\Delta,\bZ}\{ \bPhi_{\rm IPW}(\bO; \btheta) ^{\otimes 2}\} 
+ E_{Y,W,\Delta,\bZ}[\bPhi_{\rm IPW}(\bO; \btheta) \{1-\Delta/\pi_{Y,X,\bZ}(Y,W,\bZ)\}\bPsi_{\rm AIPW}(Y,\bZ; \btheta)^T]
\ese
Therefore, the efficiency condition is reduced to 
\bse
\bzero &=& E_{Y,W,\Delta,\bZ}[\{1-\Delta/\pi_{Y,X,\bZ}(Y,W,\bZ)\} \bPsi_{\rm AIPW}(Y,\bZ; \btheta) \bPhi_{\rm IPW}(\bO; \btheta)^T] \\
&& + E_{Y,W,\Delta,\bZ}[\{1-\Delta/\pi_{Y,X,\bZ}(Y,W,\bZ)\}^2 \bPsi_{\rm AIPW}(Y,\bZ; \btheta)^{\otimes 2}]  \\
&=&  E_{Y,W,\Delta,\bZ} \bigr( \{1-\Delta/\pi_{Y,X,\bZ}(Y,W,\bZ)\} \bPsi_{\rm AIPW}(Y,\bZ; \btheta) \\
&& \times [\bPhi_{\rm IPW}(\bO; \btheta) + \{1-\Delta/\pi_{Y,X,\bZ}(Y,W,\bZ)\} \bPsi_{\rm AIPW}(Y,\bZ; \btheta) ]^T \bigr) \\
&=& E_{Y,\bZ} \biggr\{ E_{W,\Delta | Y,\bZ} \bigr( \{1-\Delta/\pi_{Y,X,\bZ}(Y,W,\bZ)\} \bPsi_{\rm AIPW}(Y,\bZ; \btheta) \\
&& \times [\bPhi_{\rm IPW}(\bO; \btheta) + \{1-\Delta/\pi_{Y,X,\bZ}(Y,W,\bZ)\} \bPsi_{\rm AIPW}(Y,\bZ; \btheta) ]^T \bigr) \biggr\}.
\ese
A sufficient condition for this equation to be true is if the inner-expectation is equal to zero. More specifically: 
\bse
\bzero 
&=& \bPsi_{\rm AIPW}(y,\bz; \btheta) E_{W,\Delta | Y,\bZ} \bigr( \{1-\Delta/\pi_{Y,X,\bZ}(y,W,\bz)\} \\
&& \times [\bPhi_{\rm IPW}(\bO; \btheta) + \{1-\Delta/\pi_{Y,X,\bZ}(y,W,\bz)\} \bPsi_{\rm AIPW}(y,\bz; \btheta) ]^T \bigr) \\
&=& E_{W,\Delta | Y,\bZ} \left( \{1-\Delta/\pi_{Y,X,\bZ}(y,W,\bz)\} [\bPhi_{\rm IPW}(\bO; \btheta) + \{1-\Delta/\pi_{Y,X,\bZ}(y,W,\bz)\} \bPsi_{\rm AIPW}(y,\bz; \btheta) ]^T \right),
\ese
where the last line is obtained by pre-multiplying by $\bPsi_{\rm AIPW}(y,\bz; \btheta)^{-1}$ on both sides.

Now, we solve for $\bPsi_{\rm AIPW}(y,\bz; \btheta)$. This leads to
\bse
\bPsi_{\rm AIPW}^{\rm eff}(y,\bz; \btheta) &=& - \frac{ E_{W,\Delta | Y,\bZ} \left[ \{1-\Delta/\pi_{Y,X,\bZ}(y,W,\bz)\}\bPhi_{\rm IPW}(\bO; \btheta) \right]}{E_{W,\Delta | Y,\bZ} \left[ \{1-\Delta/\pi_{Y,X,\bZ}(y,W,\bz)\}^2 \right]} \\
&=& \frac{ E_{X | Y,\bZ} \left[ \{1-1/\pi_{Y,X,\bZ}(y,X,\bz)\} \bS_{\btheta}^F (y,X,\bz)  \right]}{E_{X | Y,\bZ}  \{1-1/\pi_{Y,X,\bZ}(y,X,\bz)\}}.
\ese
The above follows since
\bse
\bPsi_{\rm AIPW}^{\rm eff}(y,\bz; \btheta) &=& - \frac{ E_{W,\Delta | Y,\bZ} \left[ \{1-\Delta/\pi_{Y,X,\bZ}(y,W,\bz)\}\bPhi_{\rm IPW}(\bO; \btheta) \right]}{E_{W,\Delta | Y,\bZ} \left[ \{1-\Delta/\pi_{Y,X,\bZ}(y,W,\bz)\}^2 \right]} \\
&=& - \frac{ E_{W,\Delta | Y,\bZ} \left[ \{1-\Delta/\pi_{Y,X,\bZ}(y,W,\bz)\} \Delta \bS_{\btheta}^F(y,W,\bz)/\pi_{Y,X,\bZ}(y,W,\bz) \right]}{E_{W,\Delta | Y,\bZ} \{1-2 \Delta/\pi_{Y,X,\bZ}(y,W,\bz) + \Delta/\pi_{Y,X,\bZ}(y,W,\bz)^2\} } \\
&=& - \frac{ E_{X | Y,\bZ} \left[ \{1-1/\pi_{Y,X,\bZ}(y,X,\bz)\} \pi_{Y,X,\bZ}(y,X,\bz) \bS_{\btheta}^F(y,W,\bz)/\pi_{Y,X,\bZ}(y,X,\bz) \right]}{E_{X | Y,\bZ} \{1-2 \pi_{Y,X,\bZ}(y,X,\bz)/\pi_{Y,X,\bZ}(y,X,\bz) + \pi_{Y,X,\bZ}(y,X,\bz)/\pi_{Y,X,\bZ}(y,X,\bz)^2\} } \\
&=& - \frac{ E_{X | Y,\bZ} \left[ \{1-1/\pi_{Y,X,\bZ}(y,X,\bz)\} \bS_{\btheta}^F (y,X,\bz)  \right]}{E_{X | Y,\bZ}  \{-1 + 1/\pi_{Y,X,\bZ}(y,X,\bz)\}} \\
&=& \frac{ E_{X | Y,\bZ} \left[ \{1-1/\pi_{Y,X,\bZ}(y,X,\bz)\} \bS_{\btheta}^F (y,X,\bz)  \right]}{E_{X | Y,\bZ}  \{1-1/\pi_{Y,X,\bZ}(y,X,\bz)\}}.
\ese

Now, \emph{when  $\btheta$ and $\boldeta$ are unknown},  $\bPsi^{\text{eff}}_{\rm AIPW}(y, \bz; \btheta)$ continues to yield efficiency gains over the IPW estimator provided that $\btheta \independent \boldeta$. Specifically, even when $\boldeta$ is unknown and must be estimated, the same efficient augmentation component leads to efficiency gains over the IPW estimator by following a similar argument as \cite{Bartlettetal2014}. 

\subsubsection{Guaranteeing higher efficiency and reducing computational cost.}

Similar to \cite{vazquez2024establishing}, we update the augmented component with $\bPsi^{\rm updated}_{\rm AIPW}(y,\bz; \btheta)$ to ensure that it is always more efficient than the IPW estimator when only the probability $\pi_{Y,X,\bZ}(y,w,\bz)$ is correctly specified but not the augmented component $\bPsi_{\rm AIPW}^{\rm eff}(y,\bz; \btheta)$.

More specifically, we let $\bPsi^{\rm updated}_{\rm AIPW}(y,\bz) = \bLambda \bPsi_{\rm AIPW} (Y,\bZ)$ where $\bLambda$ is a $p \times q$ invertible non-random matrix and $\bPsi_{\rm AIPW} (Y,\bZ)$ is any $p$-length vector that is a function of $(Y,\bZ)$. To find the form of $\bLambda$ that will make the AIPW more efficient we again use the efficiency condition \eqref{eqn:chp3-efficiency-condition}. It follows that
\bse
\bzero &=&  \cov(\bUpsilon_{\rm AIPW} - \bUpsilon_{\rm IPW}, \bUpsilon_{\rm AIPW}) \\
&=& \cov(\bUpsilon_{\rm AIPW}) - \cov(\bUpsilon_{\rm IPW}, \bUpsilon_{\rm AIPW}) \\
&=& E_{Y,W,\Delta,\bZ}\{ \bPhi_{\rm AIPW}(\bO; \btheta) ^{\otimes 2}\} - E_{Y,W,\Delta,\bZ}\{ \bPhi_{\rm IPW}(\bO; \btheta)\bPhi_{\rm AIPW}(\bO; \btheta)^T \}.
\ese
The first part of this equation is equal to
\bse
&& E_{Y,W,\Delta,\bZ}\{ \bPhi_{\rm AIPW}(\bO; \btheta) ^{\otimes 2}\} \\
&& = E_{Y,W,\Delta,\bZ}\{ \bPhi_{\rm IPW}(\bO; \btheta) ^{\otimes 2}\} + E_{Y,W,\Delta,\bZ} [ \bPhi_{\rm IPW}(\bO; \btheta) \{1-\Delta/\pi_{Y,X,\bZ}(Y,W,\bZ)\} \bLambda \bPsi_{\rm AIPW}(Y,\bZ; \btheta)^T ] \\
&& \quad  + E_{Y,W,\Delta,\bZ}[ \{1-\Delta/\pi_{Y,X,\bZ}(Y,W,\bZ)\} \bLambda \bPsi_{\rm AIPW}(Y,\bZ; \btheta) \bPhi_{\rm IPW}(\bO; \btheta)^T ] \\
&& \quad + E_{Y,W,\Delta,\bZ} [ \{1-\Delta/\pi_{Y,X,\bZ}(Y,W,\bZ)\}^2 \bLambda \bPsi_{\rm AIPW}(Y,\bZ; \btheta)^{\otimes 2}].  
\ese
Note that in comparison to the last section, $\bLambda \bPsi_{\rm AIPW}(Y,\bZ; \btheta)$ is used as opposed to only $\bPsi_{\rm AIPW}(Y,\bZ; \btheta)$. The second part of the equation is equal to
\bse
&& E_{Y,W,\Delta,\bZ}\{ \bPhi_{\rm IPW}(\bO; \btheta)\bPhi_{\rm AIPW}(\bO; \btheta)^T \} \\
&& = E_{Y,W,\Delta,\bZ}\{ \bPhi_{\rm IPW}(\bO; \btheta) ^{\otimes 2}\} 
+ E_{Y,W,\Delta,\bZ}[\bPhi_{\rm IPW}(\bO; \btheta) \{1-\Delta/\pi_{Y,X,\bZ}(Y,W,\bZ)\}\bLambda\bPsi_{\rm AIPW}(Y,\bZ; \btheta)^T]
\ese
Therefore, our efficiency condition is reduced to 
\bse
\bzero &=& E_{Y,W,\Delta,\bZ}[\{1-\Delta/\pi_{Y,X,\bZ}(Y,W,\bZ)\} \bLambda\bPsi_{\rm AIPW}(Y,\bZ; \btheta) \bPhi_{\rm IPW}(\bO; \btheta)^T] \\
&& + E_{Y,W,\Delta,\bZ}[\{1-\Delta/\pi_{Y,X,\bZ}(Y,W,\bZ)\}^2 \bLambda\bPsi_{\rm AIPW}(Y,\bZ; \btheta)^{\otimes 2}]  \\
&=& \bLambda E_{Y,W,\Delta,\bZ}[\{1-\Delta/\pi_{Y,X,\bZ}(Y,W,\bZ)\} \bPsi_{\rm AIPW}(Y,\bZ; \btheta) \bPhi_{\rm IPW}(\bO; \btheta)^T] \\
&& + \bLambda E_{Y,W,\Delta,\bZ}[\{1-\Delta/\pi_{Y,X,\bZ}(Y,W,\bZ)\}^2 \bPsi_{\rm AIPW}(Y,\bZ; \btheta)^{\otimes 2}] \bLambda^T. 
\ese
Solving for $\bLambda$ leads to 
\bse
\bLambda_{\rm AIPW} &=& - E_{Y,W,\Delta,\bZ}[\{1-\Delta/\pi_{Y,X,\bZ}(Y,W,\bZ)\}^2 \bPsi_{\rm AIPW}(Y,\bZ; \btheta)^{\otimes 2}]^{-1} \\
&& \times E_{Y,W,\Delta,\bZ}[\{1-\Delta/\pi_{Y,X,\bZ}(Y,W,\bZ)\} \bPsi_{\rm AIPW}(Y,\bZ; \btheta) \bPhi_{\rm IPW}(\bO; \btheta)^T] 
\ese
Therefore, for any choice of $\bPsi_{\rm AIPW}(Y,\bZ; \btheta)$, we obtain a corresponding $\bLambda$ that guarantees higher efficiency than the IPW estimator. 

We let $\bPsi_{\rm AIPW}(Y,\bZ; \btheta) = E_{X|Y,\bZ} \{\bS_\btheta^F (y,X,\bZ;\btheta)\}$ for which we will denote as $\bPsi_{\rm close}(Y,\bZ; \btheta)$, as this choice has a closed form solution when we assume that $X$ follows a normal distribution and when $m(\cdot)$ is linear in $X$. See \cite{vazquez2024establishing} for the closed form solution of $E_{X|Y,\bZ} \{\bS_\btheta^F (y,X,\bZ;\btheta)\}$. Thus, our updated form is
\bse
\bLambda_{\rm AIPW} &=& - E_{Y,W,\Delta,\bZ}[\{1-\Delta/\pi_{Y,X,\bZ}(Y,W,\bZ)\}^2 \bPsi_{\rm close}(Y,\bZ; \btheta)^{\otimes 2}]^{-1} \\
&& \quad \times E_{Y,W,\Delta,\bZ}[\{1-\Delta/\pi_{Y,X,\bZ}(Y,W,\bZ)\} \bPsi_{\rm close}(Y,\bZ; \btheta) \bPhi_{\rm IPW}(\bO; \btheta)^T] 
\ese


Now, \emph{when $\btheta$ and $\boldeta$ are unknown,} we follow a similar argument as that when only $\btheta$ is unknown, except that now we find the form of $\bLambda_{\rm AIPW}$ that makes the  efficiency condition in equation (\ref{eqn:chp3-efficiency-condition}) true. 
Solving for $\bLambda_{\rm AIPW, AFT}$ leads to
\bse
\bLambda_{\rm AIPW, AFT} &=& -\biggr\{ E_{Y,W,\Delta,\bZ} \biggr( \left[ \left\{1-\Delta/\pi_{Y,X,\bZ}(Y,W,\bZ;\boldeta) \right\} \bPsi_{\rm AIPW}(Y,\bZ; \btheta) + \bA^*_\text{AIPW,AFT} \bUpsilon_\text{AFT}(\bO) \right] \\
&& \sextant \times \left\{ \Delta \bS_{\btheta}^F(Y,W,\bZ;\btheta)/\pi_{Y,X,\bZ}(Y,W, \bZ; \boldeta) + \bA^*_\text{IPW,AFT} \bUpsilon_\text{AFT}(\bO) \right\} \biggr) \biggr\}^T \\
&& \times E_{Y,W,\Delta,\bZ} \biggr( \biggr[ \left\{1- \Delta/\pi_{Y,X,\bZ}(Y,W, \bZ; \boldeta) \right\} \bPsi_{\rm AIPW}(Y,\bZ; \btheta)  + \bA^*_\text{AIPW,AFT} \bUpsilon_\text{AFT}(\bO) \biggr]^{\otimes 2} \biggr) ^{-T}.
\ese 
where 
\bse
\bA^*_\text{IPW,AFT} &=& E_{Y,W,\Delta,\bZ} \left\{   \frac{\partial}{\partial \boldeta^T}  \frac{\Delta \bS^F_\btheta(Y,W,\bZ;\btheta)}{\pi_{Y,X,\bZ}(Y,W, \bZ; \boldeta)} \right\},\\
\bA^*_\text{AIPW,AFT} &=& E_{Y,W,\Delta,\bZ} \left( \frac{\partial}{\partial \boldeta^T} \biggr[\biggr\{1- \frac{\Delta}{\pi_{Y,X,\bZ}(Y,W, \bZ; \boldeta)} \biggr\} \bPsi_{\rm AIPW}(Y, \bZ; \btheta) \biggr] \right), \text{\ and} \\
\bUpsilon_\text{AFT}(\bO) &=& E_{Y,W,\Delta,\bZ} \left\{ \frac{\partial}{\partial\boldeta^T} \bPhi_\text{AFT}(\bO; \boldeta) \right\}^{-1} \bPhi_\text{AFT}(\bO; \boldeta).
\ese 
For any choice of $\bPsi_{\rm AIPW}(y,\bz; \btheta_0)$, the construction of $\bLambda_{\rm AIPW, AFT}$ guarantees improved efficiency relative to the IPW estimator. Because $\bLambda_{\rm AIPW, AFT}$ influences only the variance of $\btheta$ and not its consistency, the same influence function can be used to characterize the asymptotic behavior of $\widehat{\btheta}$ in the presence of augmentation \cite[Theorem 6.2]{newey1994large}. In other words, the variability introduced by estimating $\widehat{\bLambda}_{\rm AIPW, AFT}$ can be ignored in the variance estimation of $\widehat{\btheta}_{\rm AIPW, AFT}$.

\section{Proof of Theorem 2: Consistency and asymptotic normality of the MLE}

\subsection{Estimation of nuisance distributions}

Similar to when we needed to estimate $f_{C|Y,\bZ}$, we can estimate $f_{X|\bZ}$ in a similar form by imposing parametric form for which it is indexed by the finite parameter set $\bgamma$. It follows that for the density $f_{X|\bZ}(c,\bz; \bgamma)$, the score equation is equal to
    \bse
     \bS_{\bgamma}(w_i,\delta_i,\bz_i; \bgamma) &=& \frac{\partial}{\partial \bgamma^T} \log f_{Y,W,\Delta,\bZ}(y_i, w_i,\delta_i, \bz_i) \\
     &\propto& \frac{\partial}{\partial \bgamma^T} \left\{ \delta_i \log f_{X|\bZ}(w_i,\bz_i;\bgamma) + (1-\delta_i) \log \int_{w\leq X} f_{ X|\bZ}( x, \bz_i;\bgamma) dx \right\} \\
     &=& \delta_i  \partial \log f_{X|\bZ}(w_i,\bz_i; \bgamma) /\partial \bgamma^T \\
     && + (1-\delta_i) \frac{\int_{w_i\leq X} \{ \partial \log f_{X|\bZ}(x,\bz_i; \bgamma) /\partial \bgamma^T  \}  f_{X|\bZ}(x,\bz_i; \bgamma) dx}{\int_{w_i\leq X} f_{X|\bZ}(x,\bz_i; \bgamma) dx} 
    \ese
    Finally, let $\wh\bgamma$ be the solution to the estimating equation 
    \be
    \label{eqn:chp3-gamma_est}
     \sumi \bPhi_{\bgamma}(\bO_i; \bgamma) = \sumi \bS_{\bgamma}(w_i,\delta_i,\bz_i; \bgamma) = \bzero.
    \ee
     Like $f_{C|Y,\bZ}$, we will assume that the density $f_{X|\bZ}$ follows a Weibull distribution based on the specification of the AFT Weibull regression model.

\subsection{Consistency}

The likelihood of the observed data can be expressed as 
\bse
&& f_{Y,W,\Delta,\bZ}(y,w,\delta,\bz) \\
&& = \left\{ \int I(w \leq c) f_{Y,X,C,\bZ}(y,w,c,\bz) dc \right\}^{\delta} \left\{ \int I(w < x) f_{Y,X,C,\bZ}(y,x,w,\bz) dx \right\}^{1-\delta} \\
&& = \left\{ \int I(w \leq c) f_{C \mid Y,X,\bZ}(c,y,w,\bz) dc f_{Y \mid X,\bZ}(y,w,\bz; \btheta) f_{X,\bZ}(w,\bz) \right\}^{\delta} \\
&& \quad \times \left\{ \int I(w < x) f_{C \mid Y,X,\bZ}(w,y,x,\bz) f_{Y\mid X,\bZ}(y,x,\bz; \btheta) f_{X,\bZ}(x,\bz) dx \right\}^{1-\delta} \\
&& \stackrel{C \independent Y | (X,\bZ)}{=} \left\{ \int I(w \leq c) f_{C \mid Y,\bZ}(c,y,\bz) dc f_{Y \mid X,\bZ}(y,w,\bz; \btheta) f_{X,\bZ}(w,\bz) \right\}^{\delta} \\
&& \quad \times \left\{ \int I(w < x) f_{C \mid Y,\bZ}(w,y,\bz) f_{Y\mid X,\bZ}(y,x,\bz; \btheta) f_{X,\bZ}(x,\bz) dx \right\}^{1-\delta} \\
&& = \left\{ \pi_{Y,X,\bZ}(y,w,\bz) f_{Y \mid X,\bZ}(y,w,\bz; \btheta) f_{X,\bZ}(w,\bz) \right\}^{\delta} \\
&& \quad \times \left\{ \int I(w < x) f_{C \mid Y,\bZ}(w,y,\bz) f_{Y\mid X,\bZ}(y,x,\bz; \btheta) f_{X,\bZ}(x,\bz) dx \right\}^{1-\delta} 
\ese
Now we take the log of likelihood which results in
\bse
\log f_{Y,W,\Delta,\bZ}(y,w,\delta,\bz) &=&  \delta \log \pi_{Y,X,\bZ}(y,w,\bz) f_{X,\bZ}(w,\bz) + \delta\log  f_{Y \mid X,\bZ}(y,w,\bz; \btheta)   \\
&& + (1-\delta) \log \left\{ \int I(w < x) f_{C \mid Y,\bZ}(w,y,\bz) f_{Y\mid X,\bZ}(y,x,\bz; \btheta) f_{X,\bZ}(x,\bz) dx \right\}
\ese
After, taking the partial derivative with respect to $\btheta$, the resulting score equation is
\bse
&& \bPhi_{\rm MLE}(\bO;\btheta) \\
&& = \delta \bS_{\btheta}^F (y,w,\bz; \btheta) + (1-\delta) \frac{\int I(w < x) f_{C \mid Y,\bZ}(w,y,\bz) \partial/\partial \btheta f_{Y\mid X,\bZ}(y,x,\bz; \btheta) f_{X,\bZ}(x,\bz) dx}{\int I(w < x) f_{C \mid Y,\bZ}(w,y,\bz) f_{Y\mid X,\bZ}(y,x,\bz; \btheta) f_{X,\bZ}(x,\bz) dx} 
\ese
The last line follows since $\partial/\partial \btheta f_{Y\mid X,\bZ}(y,x,\bz; \btheta) = \bS_{\btheta}^F (y,x,\bz; \btheta) f_{Y\mid X,\bZ}(y,x,\bz; \btheta)$. To show that the MLE is consistent, we need to show that the expectation of the score equation is equal to zero. 
It follows that
\bse
&& E_{Y,W,\Delta,\bZ} \{\bPhi_{\rm MLE}(\bO;\btheta)\} =  \int \delta \bS_{\btheta}^F (y,w,\bz) f_{Y,W,\Delta,\bZ}(y,w,\delta,\bz) dydwyd\delta d\bz \\
&& \quad + \int \left\{ (1-\delta) \frac{\int I(w < x) f_{C \mid Y,\bZ}(w,y,\bz) \bS_{\btheta}^F (y,x,\bz; \btheta) f_{Y\mid X,\bZ}(y,x,\bz; \btheta) f_{X,\bZ}(x,\bz) dx}{\int I(w < x) f_{C \mid Y,\bZ}(w,y,\bz) f_{Y\mid X,\bZ}(y,x,\bz; \btheta) f_{X,\bZ}(x,\bz) dx} \right\} \\
&& \quad \quad \times f_{Y,W,\Delta,\bZ}(y,w,\delta,\bz) dy dw d\delta d\bz \\
&& =  E_{Y,X,\bZ} \left\{ \pi_{Y,X,\bZ}(Y,X,\bZ) \bS_{\btheta}^F (Y,X,\bZ) \right\} \\
&& \quad + \int \left\{ \frac{\int I(c < x) f_{C \mid Y,\bZ}(c,y,\bz) \bS_{\btheta}^F (y,x,\bz; \btheta) f_{Y\mid X,\bZ}(y,x,\bz; \btheta) f_{X,\bZ}(x,\bz) dx}{\int I(c < x) f_{C \mid Y,\bZ}(c,y,\bz) f_{Y\mid X,\bZ}(y,x,\bz; \btheta) f_{X,\bZ}(x,\bz) dx} \right\} \\
&& \quad \quad \times \left\{ \int I(c < x) f_{C \mid Y,\bZ}(c,y,\bz) f_{Y\mid X,\bZ}(y,x,\bz; \btheta) f_{X,\bZ}(x,\bz) dx \right\} dy dc d\bz \\
&& =  E_{Y,X,\bZ} \left\{ \pi_{Y,X,\bZ}(Y,X,\bZ) \bS_{\btheta}^F (Y,X,\bZ) \right\} \\
&& \quad + \int  I(c < x) f_{C \mid Y,\bZ}(c,y,\bz) dc \ \bS_{\btheta}^F (y,x,\bz; \btheta) f_{Y\mid X,\bZ}(y,x,\bz; \btheta) f_{X,\bZ}(x,\bz) dx dy d\bz \\
&& =  E_{Y,X,\bZ} \left\{ \pi_{Y,X,\bZ}(Y,X,\bZ) \bS_{\btheta}^F (Y,X,\bZ) \right\} + E_{Y,X,\bZ} \left[ \left\{ 1-  \pi_{Y,X,\bZ}(Y,X,\bZ) \right\} \bS_{\btheta}^F (Y,X,\bZ)  \right] \\
&& =  E_{Y,X,\bZ} \left\{ \bS_{\btheta}^F (Y,X,\bZ) \right\} \\
&& =  E_{X,\bZ} [E_{Y \mid X,\bZ} \left\{ \bS_{\btheta}^F (Y,X,\bZ) \right\}] \\
&& =  E_{X,\bZ} (\bzero) = \bzero 
\ese
Since the expectation is equal to zero, then it follows that the MLE is consistent. Note that, for this result to hold we need to satisfy $\btheta \independent \balpha$---the parameters indexing $f_{C|Y,\bZ}$, need to be independent of $\btheta$.

\noindent \textbf{Robustness:} Suppose we misspecify the nuisance distribution $f_{X|\bZ}$ as  $f^*_{X|\bZ}$. The part corresponding to the CC estimator remains $E_{Y,X,\bZ} \left\{ \pi_{Y,X,\bZ}(Y,X,\bZ) \bS_{\btheta}^F (Y,X,\bZ) \right\}$. The second component is now
\bse
&& \int \left\{ (1-\delta) \frac{\int I(w < x) f_{C \mid Y,\bZ}(w,y,\bz) \bS_{\btheta}^F (y,x,\bz; \btheta) f_{Y\mid X,\bZ}(y,x,\bz; \btheta) f_{X|\bZ}^*(x,\bz) dx}{\int I(w < x) f_{C \mid Y,\bZ}(w,y,\bz) f_{Y\mid X,\bZ}(y,x,\bz; \btheta) f_{X|\bZ}^*(x,\bz) dx} \right\} \\
&& \quad \quad \times f_{Y,W,\Delta,\bZ}(y,w,\delta,\bz) dydwyd\delta d\bz \\
&& \quad + \int \left\{ \frac{\int I(c < x) f_{C \mid Y,\bZ}(c,y,\bz) \bS_{\btheta}^F (y,x,\bz; \btheta) f_{Y\mid X,\bZ}(y,x,\bz; \btheta) f_{X|\bZ}^*(x,\bz) dx}{\int I(c < x) f_{C \mid Y,\bZ}(c,y,\bz) f_{Y\mid X,\bZ}(y,x,\bz; \btheta) f_{X|\bZ}^*(x,\bz) dx} \right\} \\
&& \quad \quad \times \left\{ \int I(c < x) f_{C \mid Y,\bZ}(c,y,\bz) f_{Y\mid X,\bZ}(y,x,\bz; \btheta) f_{X|\bZ}(x,\bz) dx \right\} dy dc d\bz \\
&& \ne E_{Y,X,\bZ} \left[ \left\{ 1-  \pi_{Y,X,\bZ}(Y,X,\bZ) \right\} \bS_{\btheta}^F (Y,X,\bZ)  \right]. 
\ese
Since the component is not equal to $E_{Y,X,\bZ} \left[ \left\{ 1-  \pi_{Y,X,\bZ}(Y,X,\bZ) \right\} \bS_{\btheta}^F (Y,X,\bZ)  \right]$, then the augmentation component of the MLE does not correct for the bias present in the CC estimator. This means that we cannot guarantee that the expectation of the estimating equation for the MLE does not equal 0.  Therefore, the MLE is not robust to misspecification of $f_{X|\bZ}$. 

\subsection{Asymptotic normality}

Following a similar argument as for the AIPW estimator, the influence function of the MLE is characterized by 
\bse
\bUpsilon_{ \rm MLE, AFT} (\bO_i) &=& \bA_{\rm MLE, AFT}^{-1} \{ (\bPhi_{\rm MLE}(\bO_i; \btheta_0,\bgamma_0) \\
&& \quad \quad - E\{\partial\bPhi_{\rm MLE}(\bO;\btheta_0,\bgamma)/\partial\bgamma^T |_{\bgamma = \bgamma_0}\}[E\{\partial\bPhi_{\rm AFT}(\bO;\bgamma)/\partial\bgamma^T |_{\bgamma = \bgamma_0} \}]^{-1}\bPhi_{\rm AFT}(\bO_i;\bgamma_0) )\},
\ese
The components are 
\bse
\bA_{\rm MLE, AFT} &=& E\left\{\partial  \bPhi_{\rm MLE}(\bO; \btheta, \bgamma_0) / \partial \btheta^T |_{\btheta=\btheta_0} \right\} \\
\bB_{\rm MLE, AFT} &=&  E\{(\bPhi_{\rm MLE}(\bO; \btheta_0,\bgamma_0) \\
&& - E\{\partial\bPhi_{\rm MLE}(\bO;\btheta_0,\bgamma)/\partial\bgamma^T |_{\bgamma = \bgamma_0}\} [E\{\partial\bPhi_{\rm AFT}(\bO;\bgamma)/\partial\bgamma^T|_{\bgamma = \bgamma_0}\}]^{-1} \bPhi_{\rm AFT}(\bO;\bgamma_0) )^{\otimes2}\}.
\ese
Unlike the AIPW, $\bA_{\rm MLE, AFT}$ may not further simplified.

When the nuisance parameters indexing the nuisance distribution $f_{C|Y,\bZ}$ are assumed known, the corresponding influence function gets reduced to 
\bse
\bUpsilon_\text{MLE}(\bO_i) = - \bA_\text{MLE}^{-1} \bPhi_\text{MLE}(\bO;\bgamma_0)
\ese
where the components of the asymptotic variance are $\bA_{\rm MLE} = E\{\partial \bPhi_{\rm MLE}(\bO; \btheta, \bgamma_0)/\partial\btheta^{\trans} |_{\btheta = \btheta_0}\}$ and $\bB_{\rm MLE} = E\{\bPhi_{\rm MLE}(\bO;\btheta_0,\bgamma_0)^{\otimes2}\}$.

\section{Data Application: Survival Analysis}
\label{sec:chp3-data-application-aft}

The Weibull AFT regression model for censoring time indicated that individuals with higher baseline cUHDRS scores, longer CAG repeat lengths, and female sex had shorter expected censoring times, suggesting they exited the risk set for MCI sooner (Table~\ref{tab:chp3-hd_app_table_nuisance}). Such shorter times may reflect differential censoring, arising from earlier onset of MCI or differences in study dropout, thereby reducing the period during which censoring could occur. Education level showed no significant association with censoring time. The estimated Weibull shape parameter of 5.22 $(1/\exp(-1.652))$ indicates that the instantaneous hazard of censoring increased over the follow-up period—starting low early in follow-up and rising more sharply as the study progressed, potentially due to increasing MCI diagnoses over time.

\begin{table}[h!]
\caption{\label{tab:chp3-hd_app_table_nuisance} Weibull accelerated failure time regression parameter estimates (standard errors) of censoring time, PREDICT-HD ($n=833$).}
\centering
\resizebox{0.6\linewidth}{!}{
\begin{tabular}[t]{lccc}
\toprule
\textbf{Coefficient} & \textbf{Estimate} & \textbf{SE} &  \textbf{p-value} \\
\midrule
\addlinespace
Intercept & $4.061$ & $0.016$ & $<0.001$ \\
cUHDRS (score) & $-0.066$ & $0.005$ & $<0.001$ \\
CAG (length)  & $-0.057$ & $0.003$ & $<0.001$ \\
Education (years) & $0.003$ & $0.003$ & $0.276$ \\
Sex (female) & $-0.051$ & $0.016$ & $0.002$ \\
Log(scale) & $-1.652$ & $0.031$ & $<0.001$ \\
\bottomrule
\multicolumn{4}{l}{\textbf{Note:} cUHDRS and Education are centered at their mean;} \\
\multicolumn{4}{l}{ CAP Score is centered at its mean and scaled by a factor of 10.}
\end{tabular}}
\end{table}


\bibliographystyle{biorefs}
\bibliography{biblio}